\documentclass{article}
\usepackage{graphicx}

\begin{document}

\title{Octupolar ordering of $\Gamma_8$ ions in magnetic field}
\author{Annam\'aria Kiss and P. Fazekas\\
Research Institute for Solid State Physics and Optics\\
Budapest 114, P.O.B. 49, H-1525 Hungary.}
\maketitle
\date{}

\begin{abstract}
 We study $f$-electron lattice models which are capable of supporting 
octupolar, as well as dipolar and quadrupolar
order. Analyzing the properties of the $\Gamma_8$ ground state quartet,
we find that (111)-type combinations of the $\Gamma_5$ octupoles
${\mathcal T}_{111}^{\beta}= {\mathcal T}_x^{\beta}+{\mathcal T}_y^{\beta}
+{\mathcal T}_z^{\beta}$ are the best candidates for octupolar order
parameters. Octupolar ordering induces $\Gamma_5$-type quadrupoles as
secondary order parameter. Octupolar order is to some extent assisted, but in
its basic nature unchanged, by allowing for the presence of quadrupolar
interactions. In the absence of an external magnetic field, equivalent results
hold for antiferro-octupolar ordering on the fcc lattice. In this sense, the
choice of our model is motivated by the recent suggestion of octupolar
ordering in NpO$_2$.

The bulk of our paper is devoted to a study of the effect of an external 
magnetic field on ferro-octupolar ordering. We found that octupolar 
order survives up to a
critical magnetic field if the field is lying in specific directions, while
for general field directions, the underlying symmetry of the model is
destroyed and therefore the phase transition suppressed even in weak fields. 
Field-induced multipoles and field-induced couplings between various order 
parameters are discussed on the basis of a group theoretical analysis of 
the Helmholtz potential. We also studied the effect of octupolar ordering 
on the non-linear magnetic susceptibility which satisfies novel Ehrenfest-type 
relations at continuous octupolar transitions.

\end{abstract}


\section{Introduction}
The nature of the 25K phase transition of NpO$_2$ is a
long-standing mystery. The developments up to 1999 are reviewed in
\cite{review}. NpO$_2$ is a semiconductor with well-localized
$5f^3$ shells, thus in principle relatively easy to understand in
terms of crystal field theory and superexchange interactions.
However, it was concluded that the phase transition at 25K cannot 
be modeled by any combination of dipolar and quadrupolar
ordering phenomena. This leaves us with the possibility that the
primary order parameter is one of the octupolar moments
\cite{santini1}. Recent experimental evidence is successfully
interpreted by postulating that the primary order parameters are
$\Gamma_5$ octupoles, and that the transition is accompanied by
the induced order of $\Gamma_5$ quadrupoles
\cite{santini2,caciuffo}.

In comparison to well-studied quadrupolar phenomena, the physics of 
models supporting octupolar order is less explored 
\cite{earlier}. Motivated by recent suggestions that certain actinide and rare
earth compounds undergo octupolar ordering transitions
\cite{santini1,santini2,kuram2}, we study a lattice of $f$-shells with
$\Gamma_8$ quartet ground states, assuming symmetry-allowed
octupole--octupole and quadrupole--quadrupole interactions. Our choice of
model is motivated by certain features of NpO$_2$ (and to a lesser extent, by 
Ce$_{1-x}$La$_x$B$_6$) physics, but we could not claim that we are offering a
model for NpO$_2$.  In particular, we consider only spatially uniform
solutions though it is known that NpO$_2$ has antiferro-octupolar 
order following the four-sublattice triple-${\vec q}$ pattern 
characteristic of the fcc lattice \cite{caciuffo}. Nevertheless, our 
study of hypothetical ferro-octupolar ordering, in addition to having interest 
on its own, offers some insight into the physics of actual NpO$_2$. 
Some of the features of NpO$_2$ follow from the point group symmetry of 
the local Hilbert space, and from the mere fact of having symmetry breaking by
$\Gamma_5$-type octupolar order; we are able to account for these. 
We are, however, missing the
interesting consequences of antiferro-octupolar intersite interactions, and
${\bf Q}\ne (0,0,0)$ space group symmetry. We plan to return to these
questions in a sequel to our present work.

Analyzing the possible order parameters carried by the $\Gamma_8$ ground 
state set, we find that the best choice of the local order
parameter is the 111-type combination of $\Gamma_5$ octupoles. In
contrast to the case of CeB$_6$ (which can also be modeled as an
array of $\Gamma_8$ shells), the fourfold degeneracy is lifted in
a single continuous transition. We find that the primary ordering
of octupolar moments induces 111-type quadrupoles even in the absence 
of quadrupole--quadrupole interactions. However, allowing
for a non-vanishing $\Gamma_5$-type quadrupole--quadrupole interaction 
assists the octupolar ordering. These features are common to our 
hypothetical ferro-octupolar ordering model, and the actual 
antiferro-octupolar ordering of NpO$_2$.  This holds
in the absence of external (magnetic or strain) fields. However, our main
interest lies in the study of field effects.

The effect of external magnetic fields on ferro-octupolar transitions is 
analyzed in detail. Since octupolar ordering is a mechanism for
the spontaneous breaking of time reversal invariance, and the
application of a magnetic field removes this invariance, it might
have been guessed that the ground state degeneracy is completely
lifted in a magnetic field, and consequently no octupolar
symmetry breaking is possible. Such is indeed the case for fields
applied in general (non-symmetrical) directions: the phase
transition is smeared out by arbitrarily weak fields. However, we
found that for magnetic fields lying in certain planes, or
pointing along specific directions, there is a remaining ground
state degeneracy which is removed by a continuous
symmetry-breaking transition (or a sequence of two transitions).
The critical temperature is gradually suppressed by increasing
the field, eventually vanishing at a critical magnetic field. In
simple terms, the explanation is the following: Magnetic fields
will, in general, induce octupolar moments as a higher order
polarization effect. However, for special field directions, the
field is not able to induce at least one of the $\Gamma_5$
octupole moments. Then this moment can be generated by intersite
interaction only, and it will become non-zero below a critical
temperature. 

We stress that this latter part of our work rests on the assumption of
uniform order, and we have not attempted to to apply similar considerations to
the antiferro-octupolar phase. The essential difficulty is that an external
magnetic field will, in general, turn the octupole moments from their original
direction, and therefore the symmetrical zero-field triple-${\vec q}$ 
structure is expected to undergo a complicated distortion which is difficult 
to analyze.  

There is a great variety of multipolar moments induced by the concerted action
of pre-existing uniform octupolar order and an external magnetic field. The
existence of such polarization effects can be deduced from general symmetry
analysis (Section \ref{sec:suppression}, and particularly Sections
\ref{sec:highfield} and \ref{sec:helmholtz}); herein we follow and extend the
approach by \cite{shiina,shiina98,sakai}. Illustrative examples are provided
by mean field calculations.

\subsection{Short review of NpO$_2$}

The experimental results on NpO$_2$ are known from Refs.
\cite{review,santini1,santini2,caciuffo}. We quote only the
findings which are pertinent to our model study.

 Actinide dioxides have
the CaF$_{2}$ crystal structure at room temperature. The
sublattice of the metal ions is an fcc lattice. Np$^{4+}$ ions
have the configuration $5f^{3}$, the corresponding Hund's rule
ground state set belongs to $J=9/2$. The tenfold degenerate
free-ion manifold is split by the cubic crystal field, yielding a
$\left|\Gamma_{6}\right\rangle$ doublet and the
$\left|\Gamma_{8}^{(1)}\right\rangle$ and
$\left|\Gamma_{8}^{(2)}\right\rangle$ quartets . Neutron
spectroscopy has shown that the ground state quartet
$\left|\Gamma_{8}^{(2)}\right\rangle$ is well separated from the
first excited state (the other quartet) \cite{fournier}.

$5f^{3}$ states have both non-Kramers and Kramers degeneracy; the
latter might have been expected to be lifted by magnetic
ordering. There is, in fact, an apparently continuous phase
transition at $T_{0}=25{\rm K}$, which  was first observed as a
large $\lambda$-anomaly in the heat capacity \cite{heat}. The
linear susceptibility rises to  a small cusp at $T_{0}$, and stays
almost constant below \cite{oldsus,santini1}. The observations
were, at first, ascribed to antiferromagnetic ordering. However,
diffraction experiments failed to detect magnetic order
\cite{mannix}, and M\"ossbauer effect has put the upper bound of
$0.01\mu_{\rm B}$ on the ordered moment. Thus we can exclude
dipolar ordering, and have to consider the possibilities of
multipolar order.

A $\Gamma_8$ ground state quartet can carry a variety of local
order parameters: in addition to the $\Gamma_4$ dipoles, also
$\Gamma_3$ and $\Gamma_5$ quadrupoles, and $\Gamma_2$-,
$\Gamma_4$-, and $\Gamma_5$-type octupoles \cite{shiina}. The
nature of intersite interactions decides which of these will
actually undergo an ordering transition. Though magnetic ordering
is the commonly expected outcome, there is no rule to guarantee
that dipolar ordering is the leading instability. In fact, for
light rare earths (and also for light actinides) quadrupolar
ordering often pre-empts, or preceeds, dipolar order. However,
the possibility that octupolar order comes first, is quite novel 
\cite{kuram2}.

The possibility of an explanation in terms of quadrupolar ordering
 has been examined with great care \cite{erdos,review,caciuffo}.
The ordering pattern has to conform to the observed fact that the
$T<T_0$ phase preserves cubic symmetry, but this requirement can
be satisfied by the triple-${\vec q}$ order of $\Gamma_5$
quadrupoles \cite{santini2}. Indeed, resonant X-ray scattering
finds long-range order of the $\Gamma_{5}$ electric quadrupole
moments. However, the unquestionable appearance of quadrupolar
order cannot be the whole story. Quadrupolar ordering alone
could not resolve the Kramers degeneracy \cite{nikolaev}, and
thus there should remain low-$T$ magnetic moments giving rise to
a Curie susceptibility; this is contrary to observations.
Furthermore, muon spin relaxation shows that local magnetic
fields, with a pattern suggestive of magnetic octupoles, appear
below $T_0$ \cite{santini2}.

Though there are cases (as in CeB$_6$) when the Kramers and
non-Kramers degeneracies of the $\Gamma_8$ quartet are lifted in
separate phase transitions, this is not the case for NpO$_2$:
quadrupolar and octupolar moments appear simultaneously. This
alone suffices to show that they must be coupled, i.e., the
octupolar moments must also be of $\Gamma_5$ symmetry \cite{gamma5}. 
Still, the question may be posed whether the transition is primarily
octupolar or quadrupolar. We confirm the proposal made in
\cite{santini2}, and \cite{caciuffo}: the primary order parameter
is the octupole moment, and its ordering induces quadrupolar
moments of the same symmetry. NpO$_2$ has antiferro-octupolar order. However, 
systems with a uniform polarization of octupolar moments may, in principle, 
exist and their study is logically the first step in studying the nature of 
octupolar order, and its coupling to other kinds of order, and to external
fields. Incidentally, we find that there is a phenomenological similarity
between our findings and the observed behavior of NpO$_2$ (and also
Ce$_{1-x}$La$_x$B$_6$), the reason being that for $\Gamma_8$ Hilbert
spaces, the $\Gamma_5$ octupolar states are non-magnetic, and magnetic
susceptibility arises only due to transitions between octupolar levels. The 
mechanism relies only on the existence of an octupolar effective field, and it
is basically the same for ferro- and and antiferro-type alignments. Naturally,
the details of polarization phenomena depend on the kind of octupolar long
range order, and we hope to return to the case of antiferro-octupolar order in
a future work.  
    
In the absence of an external magnetic field, the thermodynamics of ferro- and 
antiferro-octupolar ordering is quite similar, as is also the case for more
straightforward kinds of ordering. Our $H=0$ self-consistency equation for
the uniform polarization is the same as the one for either of the sublattice
polarizations in the triple-${\vec q}$ structure. Also the results for
non-vanishing quadrupolar interactions are transferable since the coupled
quadrupolar order is of the same $\Gamma_5$ symmetry, and therefore the
four-sublattice structure remains the same. For either ferro-octupolar, or 
triple-${\vec q}$ antiferro-octupolar order we find that i) octupole--octupole 
coupling alone is sufficient to select a unique
ground state; ii) the ground state carries also quadrupolar
moment; iii) allowing for an additional quadrupole-quadrupole
interaction does not change the character of the low-$T$ phase
until the quadrupole-quadrupole interaction exceeds a threshold
value; beyond that, a purely quadrupolar transition is followed by
octupolar ordering.

\section{Octupolar moments in the $\Gamma_{8}$ quartet state}

Neutron diffraction measurements indicate that the ground state
is one of the two $\Gamma_{8}$ quartets. Since the same irrep
occurs twice, symmetry alone cannot tell us the basis functions:
their detailed form depends on the crystal field potential.
Exploiting the fact that the crystal field splittings are large,
we neglect the higher-lying levels, and describe the phase
transition within the ground state set. Since thus the sequence
and separation of levels is of little consequence, we have
arbitrarily chosen one of the quartets obtained by assuming a
purely fourth-order potential ${\cal H}_{\rm cryst}={\cal
O}_4^0+5{\cal O}_4^4$ as the ground state. We believe that this
assumption has no influence on the main features of our results
\cite{cf}. The general form of the basis states is (numerical
coefficients will be given in Appendix~\ref{appendix:a}):

\begin{eqnarray}
\Gamma_8^{1}&=&\alpha\left|\frac{7}{2}\right\rangle+\beta\left|
-\frac{1}{2}\right\rangle+
\gamma\left|-\frac{9}{2}\right\rangle\nonumber\\[2mm]
\Gamma_8^{2}&=&\gamma\left|\frac{9}{2}\right\rangle
+\beta\left|\frac{1}{2}\right\rangle+
\alpha\left|-\frac{7}{2}\right\rangle\nonumber\\[2mm]
\Gamma_8^{3}&=&\delta\left|\frac{5}{2}\right\rangle+\epsilon\left|
-\frac{3}{2}\right\rangle\nonumber\\[2mm]
\Gamma_8^{4}&=&\epsilon\left|\frac{3}{2}\right\rangle+\delta\left|
-\frac{5}{2}\right\rangle\label{eq:states}
\end{eqnarray}

It is apparent that the $\Gamma_8$ quartet is composed of two time-reversed
pairs, thus it has twofold Kramers, and also twofold non-Kramers, degeneracy.
The above choice of the basis emphasizes the presence of $J_z$ dipole
and ${\cal O}_2^0=\frac{1}{2}(3J_z^2-J(J+1))$ quadrupole moments, but of
course it is not unique. In fact, the decomposition
\begin{equation}
\Gamma_8{\otimes}\Gamma_8 = \Gamma_{1g}{\oplus}\Gamma_{4u}{\oplus}
\Gamma_{3g}{\oplus\Gamma_{5g}}{\oplus}\Gamma_{2u}{\oplus}
\Gamma_{4u}{\oplus}\Gamma_{5u}
\end{equation}
[where $g$ ($u$) indicates invariance (change of sign) under time reversal] 
shows that the subspace carries 15 different kinds of moments: three dipolar
$\Gamma_4$, two quadrupolar $\Gamma_3$, three quadrupolar $\Gamma_5$, and
seven kinds ($\Gamma_2$, $\Gamma_4$,and $\Gamma_5$) of octupolar moments
\cite{shiina}. Within the subspace (\ref{eq:states}), we can rotate so as
to get non-vanishing expectation values of any of the 15 potential order
parameters. Or in other words, if there is an effective field 
$-{\cal R}\langle {\cal R}\rangle$ (where ${\cal R}$ may be any of the fifteen
components), then the fourfold ground state
degeneracy will be at least partially lifted.

In general, intersite interactions affect all 15 moments. Lacking a
microscopic mechanism of Np--Np interactions in NpO$_2$, we have to assume
some form of the interaction, and argue from the consequences.

Restricting our attention to octupolar ordering, we have to choose between
$\Gamma_2$, $\Gamma_4$, and $\Gamma_5$. The possibility of $\Gamma_2$
octupolar ordering was first suggested by Santini and Amoretti
\cite{santini1}. In a later work, this choice was discarded because there
would be no coupling to quadrupoles \cite{santini2}. We may also add that
$\Gamma_2$ ordering would still leave us with a twofold degenerate ground
state. We can also exclude the $\Gamma_{4}$ type octupolar moments,
because symmetry allows their mixing with the magnetic dipoles, and
dipole moments are excluded by experiments.

This leaves us with the possibility of the ordering of $\Gamma_{5}$-type
octupole moments \cite{shiina}
\begin{eqnarray}
{\mathcal T}_{x}^{\beta}& = & \frac{1}{3}
(\overline{J_{x}J_{y}^{2}}- \overline{J_{z}^2J_{x}})\nonumber\\
{\mathcal T}_{y}^{\beta} & = &
\frac{1}{3}(\overline{J_{y}J_{z}^{2}}- \overline{J_{x}^2J_{y}})\nonumber\\
{\mathcal T}_{z}^{\beta} & = &
\frac{1}{3}(\overline{J_{z}J_{x}^{2}}-
\overline{J_{y}^2J_{z}})
\label{eq:tbetadef}
\end{eqnarray}
where the bars on the angular momentum operators mean the
symmetrized combinations of operators
$\overline{J_{z}J_{x}^{2}}=J_zJ_xJ_x+J_xJ_zJ_x+J_xJ_xJ_z$, etc.
Acting as a field, either of  ${\mathcal T}_{x}^{\beta}$,
${\mathcal T}_{y}^{\beta}$, or ${\mathcal T}_{z}^{\beta}$ splits
the $\Gamma_{8}$ quartet into two doublets. The doublets carry
magnetic moments. Thus assuming, say, ${\mathcal
T}_{x}^{\beta}$-type ordering would still leave us with a residual
degeneracy which should be lifted by a separate magnetic
(dipolar) phase transition.

\begin{figure}[ht]
\centering\includegraphics[totalheight=5.1cm,angle=270]{akpf_fig_1_left.ps}
\includegraphics[totalheight=5.1cm,angle=270]{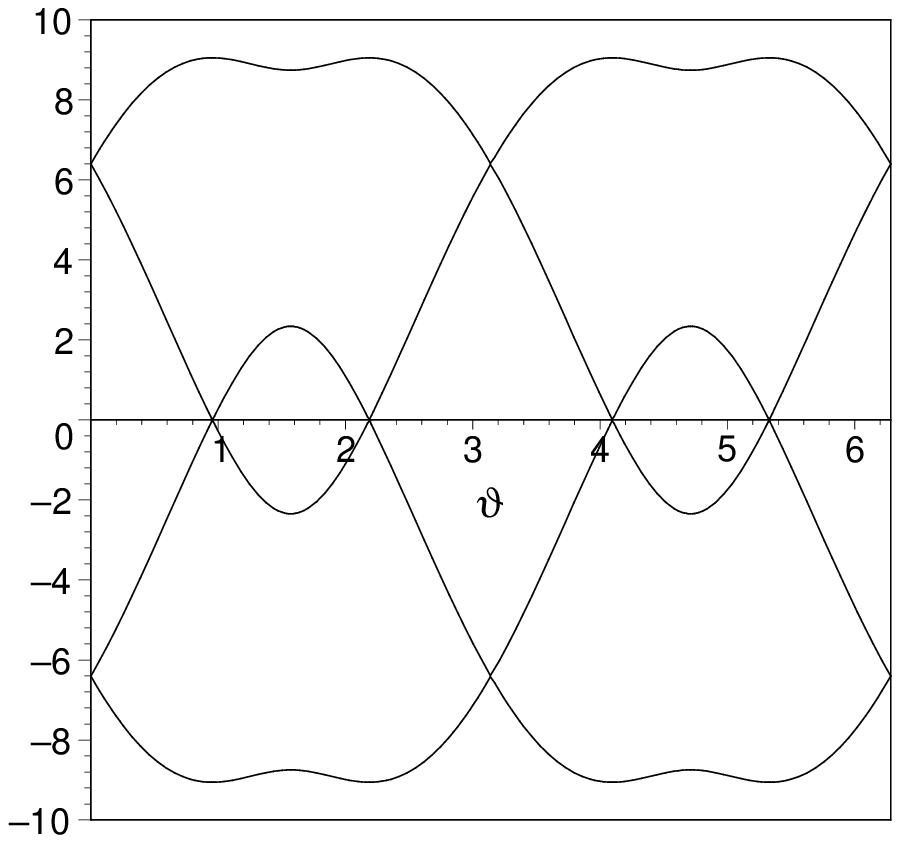}
\caption{{\sl Left}: The direction dependence of the magnitude of the 
octupolar moment ${\mathcal T}(\vartheta,\phi)$ (see equation 
(\protect{\ref{eq:tauphi}})). {\sl Right}: The spectrum of ${\mathcal
  T}(\vartheta,\phi)$ in the [1${\overline 1}$0] plane ($\phi=\pi/4$).}
\label{fig:oc1}
\end{figure}

However, we may choose a different orthogonal set of $\Gamma_5$
octupole operators as order parameters. Fig.~\ref{fig:oc1} (right) shows
the spectrum of octupoles \cite{meaning}
\begin{equation}
{\mathcal T}(\vartheta,\phi)=\sin{\vartheta}(\cos{\phi}{\mathcal
T}_x^{\beta}+ \sin{\phi}{\mathcal T}_y^{\beta})+\cos{\vartheta}\,{\mathcal
T}_z^{\beta}\label{eq:tauphi}
\end{equation}
for $\phi=\pi/4$. 
It appears that the ground state is always a singlet except for
the special points $\vartheta=0$ and $\vartheta=\pi$ (i.e., ${\mathcal
T}_z^{\beta}$ which we discussed above). Furthermore, the overall
width of the spectrum varies with $\vartheta$, reaching its maximum at
$\vartheta= \arccos{(1/\sqrt{3})}$, or equivalent positions. Thus
within the $\Gamma_8$ subspace, the three-dimensional pseudovector
of $\Gamma_5$ octupoles is ``longest'' in the (111) direction, or
in equivalent directions. At any $(\vartheta,\phi)$ the maximum of the
absolute value of the eigenvalues is
taken as the magnitude of the octupolar moment (note that the spectrum is
symmetrical about 0). This is the quantity shown in Fig.~\ref{fig:oc1}
(left). We see sixteen maxima. However, it is easy to check that the plotrange
for $(\vartheta,\phi)$ gives {\sl two} points for every direction, therefore
the number of maxima is only eight. The directions (111), $({\overline 1}11)$, 
$(1{\overline 1}1)$, and $(11{\overline 1})$ are equivalent by cubic symmetry,
and for each direction, the octupole moment may be of either sign
(Fig.~\ref{fig:oc1} (right)).    

Thinking of the ordering as caused by
an octupole-octupole interaction with cubic symmetry \cite{cubic}
\begin{equation}
{\cal H}_{\rm oc} = -J_{\rm oc} \sum_{i,j} ({\mathcal T}_{i,x}^{\beta}
{\mathcal T}_{j,x}^{\beta}+{\mathcal T}_{i,y}^{\beta}
{\mathcal T}_{j,y}^{\beta}+{\mathcal T}_{i,z}^{\beta}
{\mathcal T}_{j,z}^{\beta})
\label{eq:ocham}
\end{equation}
it is plausible that it will occur in one of the (111) directions. We may
think of it as the cubic crystal field giving rise to an octupolar single-ion
anisotropy with the (111) directions as easy axes. Thus our candidates for
order parameters are
\begin{eqnarray}
{\mathcal T}_{111}^{\beta}&=& {\mathcal T}_x^{\beta}+
{\mathcal T}_y^{\beta}+{\mathcal T}_z^{\beta}\nonumber\\[2mm]
{\mathcal T}_{{\overline 1}11}^{\beta}&=& {\mathcal T}_x^{\beta}-
{\mathcal T}_y^{\beta}-{\mathcal T}_z^{\beta}\nonumber\\[2mm]
{\mathcal T}_{1{\overline 1}1}^{\beta}&=&
-{\mathcal T}_x^{\beta}+{\mathcal T}_y^{\beta}-
{\mathcal T}_z^{\beta}\nonumber\\[2mm]
{\mathcal T}_{11{\overline 1}}^{\beta}&=&
-{\mathcal T}_x^{\beta}-
{\mathcal T}_y^{\beta}+{\mathcal T}_z^{\beta}\, .
\label{eq:111states}
\end{eqnarray}
The four minima seen in Fig. \ref{fig:oc1} belong to $\pm
{\mathcal T}_{111}^{\beta}$ and $\pm {\mathcal T}_{11{\overline
1}}^{\beta}$ \cite{scheme}. We have checked that the ground state
of either of these operators carries a $\Gamma_{5}$-type
quadrupolar moment, but {\sl no magnetic dipole moment}. 

We note that these single-ion properties would be useful for modelling
NpO$_2$; however, the interionic interactions may still be chosen as either 
ferro-octupolar or antiferro-octupolar. Let us furthermore point it out  
that the four equivalent states (\ref{eq:111states}) are ideally suited 
for constructing a four-sublattice ground state for nearest-neighbor 
antiferro-octupolar coupling on an fcc lattice. This would correspond 
to the experimentally motivated suggestion of
triple-$\overrightarrow{q}$ order by Caciuffo et al. \cite{caciuffo}. 
The triple-$\overrightarrow{q}$ octupolar
ordering can then induce the observed triple-$\overrightarrow{q}$
structure of the $\Gamma_{5}$ quadrupoles 
\begin{equation}
{\mathcal O}_{111} = {\cal O}_{xy}+{\cal O}_{yz}+{\cal O}_{zx}\, ,
\label{eq:o111}
\end{equation}
etc., as a secondary order parameter. In the absence of a magnetic field,
our mean field results are formally valid for either the ferro- or
antiferro-octupolar case, while as far as field effects
are concerned, we stick definitely to the former case. We have to refrain
from making detailed comments on NpO$_2$ until we completed work on the
magnetic properties. 

Denoting the ground state of ${\mathcal T}_{111}$ by
$|\phi_0\rangle$, we quote the numerical values from Appendix~\ref{appendix:b}:
\begin{eqnarray}
\left\langle\phi_{0}\left|{\mathcal
T}_{111}^{\beta}\right|\phi_{0}\right\rangle & = & A = -15.683\nonumber \\
\left\langle\phi_{0}\left|{\mathcal
O}_{111}\right|\phi_{0}\right\rangle & = & B = 8.019
\nonumber\\
\left\langle\phi_{0}\left|J_{x}\right|\phi_{0}\right\rangle & = &
\left\langle\phi_{0}\left|J_{y}\right|\phi_{0}\right\rangle=
\left\langle\phi_{0}\left|J_{z}\right|\phi_{0}\right\rangle=0
\label{eq:numbers}
\end{eqnarray}
and
\begin{equation}
 \left\langle\phi_{0}\left|{\cal
O}_{xy}\right|\phi_{0}\right\rangle =
\left\langle\phi_{0}\left|{\cal
O}_{yz}\right|\phi_{0}\right\rangle =
\left\langle\phi_{0}\left|{\cal
O}_{xz}\right|\phi_{0}\right\rangle\, .
\label{eq:os}
\end{equation}

One may be wondering whether the $\Gamma_5$ quadrupoles by
themselves would like a different orientation than the one forced
upon them by the octupoles. This is not the case: a calculation
shows that the length of the pseudovector $({\cal O}_{xy}, {\cal
O}_{yz},{\cal O}_{zx})$ is the same in all directions. There is
no single-ion anisotropy for the $\Gamma_5$ quadrupoles; picking
the (111) solution is exclusively the octupoles' doing.

\section{The octupolar--quadrupolar model}

We assume the presence of $\Gamma_5$-type quadrupolar and octupolar
interactions
\begin{equation}
{\cal H} = {\cal H}_{\rm oc} + {\cal H}_{\rm quad}
\label{eq:ham}
\end{equation}
where ${\cal H}_{\rm oc}$ was given in (\ref{eq:ocham}) and analogously
\begin{equation}
{\cal H}_{\rm quad} = -J_{\rm quad} \sum_{i,j} ({\mathcal
O}_{i,xy} {\mathcal O}_{j,xy}+{\mathcal O}_{i,yz} {\mathcal
O}_{j,yz}+{\mathcal O}_{i,zx} {\mathcal O}_{j,zx})\, .
\label{eq:quadham}
\end{equation}
For the sake of simplicity, we assume $J_{\rm quad}/J_{\rm oc}\ge
0$.

For ferro-octupolar coupling ($J_{\rm oc}>0$), we may assume uniform (111)
order. The mean-field single-site Hamiltonian is of the form 
\begin{equation}
{\cal H}_{\rm MF} = -{\mathcal T}_{111}^{\beta}
\langle{\mathcal T}_{111}^{\beta}\rangle -j{\mathcal O}_{111}
\langle{\mathcal O}_{111}\rangle
\label{eq:effham0}
\end{equation}
where we have chosen the octupolar effective field amplitude as the energy
unit, and $j=J_{\rm quad}/J_{\rm oc}$.  Henceforth we assume that  
all effects arising from lattice geometry, and the detailed form of the 
interactions are included in $J_{\rm oc}$, $J_{\rm quad}$, and hence also in
$j$. 

We note that formally the same mean field problem arises by assuming
antiferro-octupolar interactions, and postulating four sublattices with
local order parameters as defined in (\ref{eq:111states}) and the analogous
quadrupolar moments.

The temperature dependence of the order parameters ${\mathcal
T}=\langle{\mathcal T}_{111}^{\beta}\rangle$ and
$q=\left\langle{\mathcal O}_{111}\right\rangle$ is obtained by
the numerical solution of the self-consistency equations derived
from diagonalizing (\ref{eq:effham0}) in the basis
(\ref{eq:states}). The overall behavior is similar to that found
in dipolar--quadrupolar models used in the description of Pr
compounds \cite{lib,amk}; however, now octupoles play the role of dipoles.

The dimensionless free energy belonging to (\ref{eq:effham0}) is
\begin{equation}
{\cal F} = \frac{1}{2}{\mathcal T}^2
+\frac{1}{2} j q^2
 -t\,{\rm ln}(2{\rm exp}(-Bjq/t){\rm cosh}(A{\mathcal T}/t) +
2\,{\rm exp}(Bjq/t)) \label{eq:oqf}
\end{equation}
where $A$ and $B$ were introduced in (\ref{eq:numbers}), and 
$t=k_{\rm B}T/J_{\rm oc}$ is the dimensionless temperature.

Octupolar order (${\mathcal T}\not=0$) induces quadrupolar moment
even in the absence of a quadrupolar coupling, as we can see from
setting $j=0$ in ${\partial {\cal F}}/{\partial q}=0$
\begin{eqnarray}
q=B\frac{{\rm exp}(A{\mathcal T}/t)+{\rm exp}(-A{\mathcal T}/t)-2} 
{{\rm exp}(A{\mathcal T}/t)+{\rm exp}(-A{\mathcal T}/t)+2}\, . 
\label{eq:q}
\end{eqnarray} 
The $T\to 0$ limit is expressed in (\ref{eq:numbers}). It states
that by construction, the (111)-type octupolar eigenstates carry
(111)-type quadrupolar moments. The same state of affairs
prevails as long as ${\mathcal T}\ne 0$. In the ``para'' phase
above the transition temperature, all moments vanish.

\begin{figure}
\centering
\includegraphics[totalheight=6cm,angle=270]{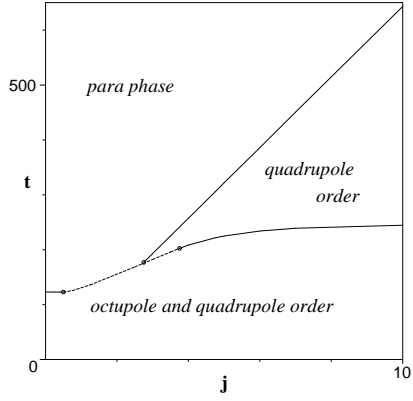}
\vspace{5mm}
\caption{The mean field phase diagram of the zero-field
quadrupolar-octupolar model (\protect\ref{eq:ham}) in the quadrupolar
coupling--temperature plane ($j=J_{\rm quad}/J_{\rm oc}$ and 
$t=Tk_{\rm B}/J_{\rm oc})$). The dashed and continuous 
lines signify first and second order phase transitions, respectively.
Observe the regime of first-order transitions bounded by two tricritical
points (marked by black dots).}\label{fig:oqphased}
\end{figure}

The continuous phase transitions of the model (\ref{eq:ham}) can be described
by the Landau expansion of the free energy (\ref{eq:oqf})
\begin{eqnarray}
{\cal F}&\approx & {\cal F}_{0}+ 
\left(\frac{1}{2}-\frac{A^2}{4t}\right){\mathcal T}^2
+\left(\frac{j}{2}-\frac{B^2j^2}{2t}\right)q^2
+ \frac{1}{4}\frac{BjA^2}{t^2}q{\mathcal T}^2 + 
\frac{A^4}{96t^3}{\mathcal T}^4
\nonumber
\\[2mm]
& & - \frac{BjA^4}{24t^4}q{\mathcal T}^4 
+ \frac{B^4j^4}{12t^3}q^4 - 
\frac{B^3j^3A^2}{12t^4}q^3{\mathcal T}^2 + ...
 \label{eq:lappr}
\end{eqnarray}
where ${\cal F}_{0}$ is the non-critical part of the free energy.

Critical temperatures are defined by the change of sign in the coefficient 
of either of the quadratic terms. Upon lowering the temperature, for small 
$j$, mixed octupolar--quadrupolar, while for large $j$, pure quadrupolar 
order sets in first. At intermediate $j$, there is a regime of first order
transitions (Fig.~\ref{fig:oqphased}). 

We consider first the weak-$j$ limit. The critical temperature is 
$t_{\rm oc}=A^2/2$. At $t_{\rm oc}$, octupolar moment appears as the primary
order parameter, but there is also induced quadrupolar order. Minimizing
${\cal F}$ with respect to $q$, we get 
\begin{equation}
q = \frac{BA^2{\cal T}^2}{4t(jB^2-t)}\, .
 \end{equation}
Thus terms of order $q^2$ and $q{\cal T}^2$ are effectively of $O({\cal
  T}^4)$. Minimizing with respect to ${\cal T}$, the critical behavior
of the octupolar and quadrupolar moment 
\begin{eqnarray}
{\cal T} & \approx &
\sqrt{\frac{A^2}{2}-t}\;\sqrt{\frac{6(A^2-2B^2j)}{A^2-8B^2j}}
\label{eq:occrit}
\\[2mm]
 q & \approx & - \frac{6B}{A^2-8B^2j} \; \left(\frac{A^2}{2}-t\right)
\label{eq:qcrit}
\end{eqnarray}
is characteristic of the mean field solution for primary, and secondary, 
order parameters.

\begin{figure}
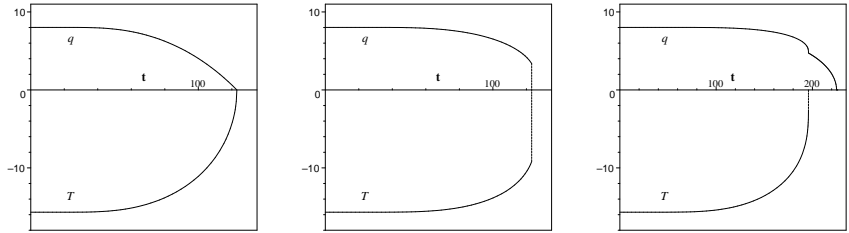

\centering
\includegraphics[totalheight=3.8cm,angle=270]{akpf_fig_3_left.ps}
\includegraphics[totalheight=3.8cm,angle=270]{akpf_fig_3_middle.ps}
\includegraphics[totalheight=3.8cm,angle=270]{akpf_fig_3_right.ps}
\caption{Octupolar (${\cal T}$) and quadrupolar ($q$) order parameters as
a function of $t=k_{\rm B}T/J_{\rm oc}$ for $J_{\rm quad}/J_{\rm oc}=0$  
({\sl left}), $J_{\rm quad}/J_{\rm oc}=0.75$ ({\sl center}), 
and $J_{\rm quad}/J_{\rm oc}=3.5$ ({\sl right}). }\label{fig:oqop}
\end{figure}

The coefficient 
\[
\frac{A^4}{96t^3}\frac{4B^2j-t}{B^2J-t}
\]
of the combined fourth order $O({\cal T}^4)$ term of ${\cal F}$ 
changes sign at $t=4B^2j$. Equating this
with the critical temperature, we identify the coordinates of the lower
tricritical point as $j_{{\rm tri},1}=A^2/8B^2=0.48$, and  
$t_{{\rm tri},1}\approx 123$. The critical temperature is constant for 
$j\le j_{{\rm tri},1}$. For $j$ exceeding $j_{{\rm tri},1}$ the 
transition becomes first order. The nature of the coupled orders 
does not change, but they become more stabilized, and the common 
transition sets in at higher temperatures (Fig.~\ref{fig:oqphased}). 
However, the ground state moments remain independent of the coupling 
strengths: $q_{T\to 0}=B$, and ${\mathcal T}_{T\to 0}=A$. 
Representative temperature dependences
of ${\mathcal T}$ and $q$ are shown in Fig.~\ref{fig:oqop}. 

At large $j$, the first instability is associated with the change of sign of
the coefficient of the $q^2$ term: pure quadrupolar order sets in at
$t_{\rm quad}=B^2j$. This critical line meets the boundary of first-order
transitions at the critical end point $j_{\rm end}\approx 2.75$, 
$t_{\rm end}\approx 177$ (Fig.~\ref{fig:oqphased}). For $j>j_{\rm end}$ 
there are two phase transitions: the onset of pure quadrupolar order is
followed by the emergence of mixed octupolar--quadrupolar order at 
$t_{\rm oc}$. The lower phase transition is of first order up to the second
tricritical point $j_{{\rm tri},2}\approx 3.75$, 
$t_{{\rm tri},2}\approx 185$. For $j<j_{{\rm tri},2}$, the onset 
of octupolar order is reflected in a discontinuity of $q$ 
(Fig.~\ref{fig:oqop}, right). For $j>j_{{\rm tri},2}$, both transitions are
continuous.    

 Deep inside the
quadrupolar ordered phase, the development of the octupolar order
is essentially unaffected by what the quadrupoles are doing,
apart from a weak effect on the transition temperature $t_{\rm oc}$
(note in Fig.~\ref{fig:oqphased} that $t_{\rm oc}$ saturates to a
constant). Though in this regime, we cannot use Landau expansion to determine
$q$, we may assume that it is near its ground state value $B$,
and use a low-order expansion in ${\mathcal T}$ to obtain in the
large-$j$ limit
\begin{eqnarray}
\lim_{j\to\infty} t_{\rm oc}=\lim_{j\to\infty}
\frac{A^2{\rm exp}(Bqj/t_{\rm oc})}{{\rm exp}
(Bqj/t_{\rm oc})+{\rm exp}(-Bqj/t_{\rm oc})}= A^2 \approx 246\, .
\label{eq:highjlim}
\end{eqnarray}

In familiar phase diagrams of dipolar--quadrupolar models, the
mixed order would be completely suppressed at $J_{\rm
quad}/J_{\rm dipole}\to\infty$ (see, e.g., Fig. 3 of Ref.~\cite{amk}). 
In contrast, we find the finite saturation value (\ref{eq:highjlim}) as
$J_{\rm quad}/J_{\rm oc}\to\infty$. 
The peculiarity of the situation depicted in Fig.~\ref{fig:oqphased}
is the endurance of octupolar order even with infinitely strong
quadrupolar coupling. The reason, as we understood earlier, is
that in the $\Gamma_8$ subspace the $\Gamma_5$ quadrupoles are
completely isotropic, thus they can accomodate a
reorientation of the basis states without sacrificing any of
their rigid order.

Even confining our attention to uniform states, the effects of an external
magnetic field are variegated: It may gradually suppress octupolar ordering,
without changing its character (111 direction); it may split the transition in
two (001 direction); it may change the character of octupolar order but still
facilitate a phase transition (11c direction); or it may completely forbid
octupolar ordering (non-symmetrical directions). We will understand this 
in detail in Sections \ref{sec:suppression} and \ref{sec:symmetry}. 
A straightforward characterization
of field effects in terms of the magnetization curve and its derivatives (the
susceptibilities) is possible in the ${\bf H}\parallel (111)$ case 
only. This is the subject of the next Subsection. 

We emphasize that our entire analysis of magnetic field effects is confined to
spatially uniform states, and does not cover the cases of supercell ordering,
such as the experimentally observable antiferro-octupolar order of NpO$_2$.
The basic difficulty is that the inter-sublattice angles of various moments 
may get changed by the field; this effect will be treated in a subsequent 
work.

\subsection{Non-linear susceptibility: the ${\bf H}\parallel (111)$ case}
\label{sec:susc}

\begin{figure}
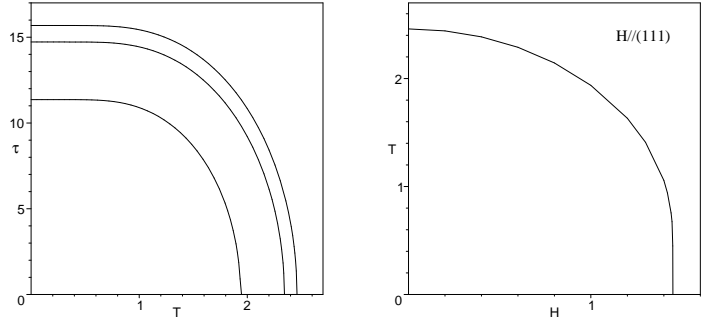

\centering
\includegraphics[totalheight=4.9cm,angle=270]{akpf_fig_4_left_nn.ps}
\includegraphics[totalheight=4.9cm,angle=270]{akpf_fig_4_right_nn.ps}
\caption{{\sl Left}: The $\langle{\mathcal T}_{111}^{\beta}\rangle={\cal T}$ 
octupolar order parameter as a function of the temperature 
for $H=0$, 0.5, and 1.0 
(${\bf H}\parallel(111)$, $H$ in units of $g\mu_{\rm B}$). 
{\sl Right}: the $T-H$ phase diagram of the
$J_{\rm oc}=0.02k_{\rm B}$, $J_{\rm quad}=0$ model for $H$ pointing 
in the (111) direction. The transition is continuous all along the phase
boundary.}\label{fig:htpd}
\end{figure}

In certain symmetry
directions such as (111), the magnetic field merely acts to
suppress uniform octupolar ordering gradually (Fig.~\ref{fig:htpd}, left). 
It appears that the phase boundaries can be scaled onto a common 
curve by introducing the field-dependent transition temperature
\begin{equation}
\frac{T_{\rm oc}(H)}{T_{\rm oc}(H=0)} \approx 1 -a_H\cdot H^2-b_H
\cdot H^4 \ldots
\label{eq:TcrH}
\end{equation}

This
bears some similarity to the field-induced suppression of
antiferro-quadrupolar order in PrFe$_4$P$_{12}$ \cite{amk}.

Our starting point is the mean-field-decoupled hamiltonian
\begin{eqnarray}
{\cal H} & = & {\cal H}_{\rm oc} + {\cal H}_{\rm quad} + {\cal
H}_{\rm Z} = {\cal H}_{\rm oc} + {\cal H}_{\rm quad} - {\bf
H}{\cdot}{\bf J} 
\nonumber\\[1mm]
& = & -{\mathcal T}\cdot{\mathcal T}_{111}^{\beta} -j q {\mathcal O}_{111} 
- HJ_{111} 
\label{eq:zqham}
\end{eqnarray}
where the notations follow (\ref{eq:effham0}), $J_{111}=(J_x+J_y+J_z)$, and 
in the Zeeman term  $H$ is the reduced magnetic field.

Multipolar phase transitions, even when non-magnetic, tend to
have a strong signature in the non-linear magnetic response. The
case of quadrupolar transitions has been extensively studied \cite{morin}. 
To obtain analogous results, we expand the free energy corresponding to
(\ref{eq:zqham})
\begin{eqnarray}
{\cal F}({\cal T},q,H) & = & 
  \frac{1}{2}{\cal T}^2 + \frac{j}{2} q^2 \nonumber\\[1mm]
& & - 
    t\cdot \ln\; [ 2\exp{(-Bjq/t)} 
\cosh{\left(\sqrt{g_H^2 H^2+ A^2{\cal T}^2}/t\right)}  \nonumber\\[2mm]
& &  \mbox{\ \ \ \ \  \ \ \ \ }+  2 \exp{(Bjq/t)}\cosh{(y_H H/t)} ]\, .
\label{eq:magnland1}
\end{eqnarray}
Here $g_H$ and $y_H$ are the two parameters of the Zeeman splitting scheme of
the $\Gamma_8$ subspace (Appendix~\ref{appendix:a}). 
The overall shape of the phase boundary in the
$t$--$H$ plane is obtained by expanding ${\cal F}({\cal T},q,H)$ 
in powers of ${\cal T}$, and identifying the coefficient of the 
${\cal T}^2$-term
\begin{equation}
c_2(H,t)=\frac{1}{2}-\frac{A^2}{2 g_H H}{\cdot}
\frac{\sinh{(g_H H/t)}}{\cosh{(g_H H/t)}+ \cosh{(y_H H/t)}}\, .
\label{eq:alfa}
\end{equation}
Solving $c_2(H,t)=0$ gives a line of continuous transitions in
the $t$--$H$ plane (Fig.~\ref{fig:htpd}, right).

It is interesting that the  octupole ordered phase can be suppressed 
gradually by a magnetic field (Fig~\ref{fig:htpd}, left). Dipole 
ordering and octupole ordering are two
independent ways to break time reversal invariance. However, octupolar
moments are due to currents with zero total circulation, while dipole moments
arise from non-zero integrated circulation. In a finite field 
${\bf H}\parallel(111)$, the $5f^3$ ion
must be able to sustain both kinds of currents simultaneously.

${\cal F}({\cal T},q,H)$ has to be expanded to 
$O(H^4)$ in order to derive both the susceptibility $\chi$, and the 
non-linear magnetic susceptibility $\chi_3$. We do not give the detailed
formulas here, but discuss the terms giving rise a quadratic shift of the
transition temperature in (\ref{eq:TcrH})
\begin{eqnarray}
{\cal F}({\cal T},q,H) & \approx & {\cal F}({\cal T},q,H=0) \nonumber\\[1mm]
& & + \frac{g_H^2 - y_H^2}{4t^2} Bjq H^2 -
\frac{g_H^2}{12t^4} H^2 A^2Bjq{\cal T}^2 \nonumber\\[1mm]
& & +  \frac{g_H^2 + 3y_H^2}{48t^3} 
A^2 {\cal T}^2 H^2 \, .
\label{eq:magnland2}
\end{eqnarray}
The first term in the second line describes field-induced $\Gamma_5$ 
quadrupoles. The general nature of octupoles would allow the presence of a 
${\cal T}H$ term; it is the peculiarity of the (111) direction that it does
not appear here. We have omitted field-induced terms of the non-critical part
of the free energy; they have to be included when calculating the
susceptibilities.  

Further calculation is analogous to that given for the $H=0$ case. Minimizing
with respect to $q$ gives for the secondary order parameter 
\begin{equation}
q \approx -\frac{B}{4t(t-B^2j)}\cdot \left( A^2{\cal T}^2 + 
(g_H^2-y_H^2) H^2\right) \, .
 \label{eq:magnland3}
\end{equation}
Replacing this back into (\ref{eq:magnland2}), we can determine the optimum
value of the primary order parameter ${\cal T}$. Here we quote only the result
for the quadratic shift of the transition temperature (cf. Equation 
(\ref{eq:TcrH}))
\begin{equation}
a_H = \frac{1}{6A^4(A^2-2B^2j)} \left[ (A^2-8B^2j) g_H^2 + 3A^2 y_H^2
\right]\, .
 \label{eq:magnland4}
\end{equation}
One of the contributions to $a_H$ vanishes at the tricritical point ($j\to 
A^2/8B^2$), while the other remains finite.

\begin{figure}
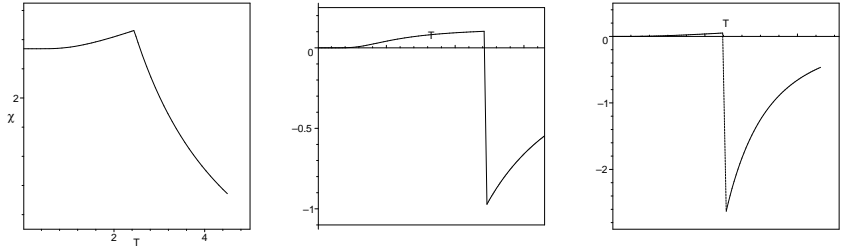

\centering
\includegraphics[totalheight=3.8cm,angle=270]{akpf_fig_5_left_nn.ps}
\includegraphics[totalheight=3.8cm,angle=270]{akpf_fig_5_middle_nn.ps}
\includegraphics[totalheight=3.8cm,angle=270]{akpf_fig_5_right_nn.ps}
\caption{Linear susceptibility ({\sl left}), temperature
derivative of linear susceptibility ({\sl center}) and nonlinear
susceptibility ({\sl right}) as a function of temperature for
magnetic field parallel to (111) direction ($J_{\rm oc}=0.02k_{\rm B}$, 
$J_{\rm quad}=0$).}\label{fig:osus}
\end{figure}

Representative results for the linear
susceptibility $\chi=-{\partial^2 {\cal F}}/{\partial H^2}$, and
the third-order susceptibility $\chi_3=-{\partial^4 {\cal
F}}/{\partial H^4}$,  are shown in Fig.~\ref{fig:osus}. 
The octupolar transition appears as a cusp in $\chi$ (Fig.~\ref{fig:osus}, 
left). The cusp can be also represented as the discontinuity
of ${\partial\chi}/{\partial T}$ (Fig.~\ref{fig:osus}, middle).
The non-linear susceptibility has a discontinuity from positive
to negative values (Fig.~\ref{fig:osus}, right). These anomalies are related
to each other, and the specific heat discontinuity  $\Delta C$, via  the 
Ehrenfest-type equation \cite{ehrenfest1}
\begin{equation}
\frac{a_H}{T}\Delta C+\frac{1}{12a_H}\Delta {\chi}_{3}=
\Delta\left(\frac{\partial {\chi}}{\partial
T}\right)\, .
\label{eq:ef}
\end{equation}
The derivation
\cite{ehrenfest2} of (\ref{eq:ef}) relies only on the fulfillment of
(\ref{eq:TcrH}) to order $H^3 \cite{ehrenfest3}$.
In particular, $b_H$ does not come
into (\ref{eq:ef}). The relationship (\ref{eq:TcrH}) is often found for the
critical temperature of transitions to non-ferromagnetic phases like
antiferromagnets, spin-gapped phases, quadrupolar order, etc. Octupolar
ordering belongs to this class of transitions.

Though the example shown in Fig.~\ref{fig:osus} was for $J_{\rm quad}=0$, the
relationship (\ref{eq:ef}) holds everywhere along the lines of continuous
phase transitions shown in Fig.~\ref{fig:oqphased}. As long as we are dealing 
with ordinary second order transitions, Landau theory would be consistent with
all the discontinuities appearing in (\ref{eq:ef}) being finite. 
However, approaching a tricritical point $\Delta C\to\infty$, and
(\ref{eq:ef}) allows several scenarios. We note from (\ref{eq:magnland4}) that
generically $y_H\ne 0$, thus $a_H$ remains finite, and then the 
simplest expectation is that
the divergence of $\Delta C$ is matched by that of 
$\Delta {\partial {\chi}}/{\partial T}$. Such is indeed the finding for our
standard $\Gamma_8$ subspace specified in Appendix~\ref{appendix:a}. The same
holds for $\Gamma_8$ subspaces derived from a combination of fourth-order and
sixth-order crystal field potentials. However, at one particular value of the  
ratio of the sixth-order and fourth-order terms, the Zeeman spectrum 
consists of a doublet and two singlets, i.e., the $y_H=0$ case is realized. 
For this special model \cite{solterdos}   
$a_H\to 0$ as one approaches the
tricritical point, and the field dependence of the transition temperature is
purely quartic (\ref{eq:TcrH}). Here, a peculiar form of the Ehrenfest 
relation can be derived. $a_H\to 0$ cancels the mean-field divergence of
$\Delta C$, and  at the same time $\Delta{\chi}_3\to 0$. The
discontinuity of $\left({\partial {\chi}}/{\partial T}\right)$ is now 
balanced by that of the fifth-order non-linear susceptibility
\begin{equation}
\frac{1}{120}\Delta\chi_5 = b_H\, \Delta{\left(\frac{\partial {\chi}}{\partial
T}\right)}.
\label{eq:chi5}
\end{equation}
It should be interesting to find a situation where (\ref{eq:chi5}) is 
experimentally testable \cite{nonlin}.

\section{Suppression of ferro-octupolar order by magnetic field}
\label{sec:suppression}

Next, we consider the effect of a finite magnetic field of arbitrary
orientation on a system of interacting $\Gamma_5$ octupoles. Henceforth, 
our mean field arguments will be based on a simplified version of 
(\ref{eq:zqham})
\begin{equation}
{\cal H} = {\cal H}_{\rm oc} + {\cal H}_{\rm Z} = {\cal H}_{\rm oc}-
{\bf H}{\cdot}{\bf J}
\label{eq:zham}
\end{equation}
where ${\cal H}_{\rm oc}$ is taken from (\ref{eq:ocham}), and $J_{\rm
  quad}=0$. Here, as in Section \ref{sec:symmetry}, we will confine our
  attention to uniform states. An essential extension of our argument 
would be needed to cover the case of NpO$_2$. 

\begin{figure}[ht]
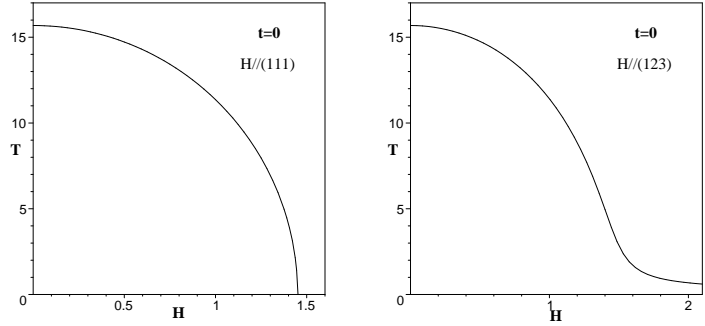

\centering
\includegraphics[totalheight=4.9cm,angle=270]{akpf_fig_6_left.ps}
\includegraphics[totalheight=4.9cm,angle=270]{akpf_fig_6_right.ps}
\caption{{\sl Left}: Field-induced ground state transition from
octupolar order to the disordered state for ${\bf H}\parallel(111)$.
{\sl Right}: Expectation value of the octupole moment in the ground
state as a function of magnetic field for ${\bf H}\parallel(123)$. For
fields pointing in non-symmetric directions, there is no sharp
phase transition.} \label{fig:fields}
\end{figure}

First, let us reconsider the case of a field ${\bf H}\parallel(111)$
(Fig.~\ref{fig:htpd}). The nature of the octupolar order
parameter is not influenced by the field, only its saturation
value, and the transition temperature, are scaled down.
Conversely: since there is a $T=0$ phase transition at a critical field 
$H_{\rm cr}$
(Fig.~\ref{fig:fields}, left), there must exist (in mean field
theory) a finite-$T$ ordered phase at $H<H_{\rm cr}$ 
(Fig.~\ref{fig:htpd}, left). Generally,
the nature of the field direction dependence of octupolar ordering
can be studied by confining our attention to the ground state
($T=0$). First, we use mean field theory; later, we give 
general symmetry arguments.

Let $E_0$ be the minimal eigenvalue of the mean field hamiltonian
which contains both the external magnetic field, and the
octupolar effective field
\begin{eqnarray}
E_{0}(\langle {\mathcal T}_{\bf H}\rangle)=\left\langle {\mathcal
H}\right\rangle= -\left\langle\Phi_{0}\left| {\mathcal T}_{\bf
H}\right|\Phi_{0}\right\rangle \left\langle\ {\mathcal T}_{\bf
H}\right\rangle-{\bf H}{\cdot}\left\langle {\bf J} \right\rangle
\label{eq:en0}
\end{eqnarray}
where $|\Phi_{0}\rangle$ is the interacting ground state, and 
$\langle\,\rangle$ denotes expectation values taken with 
$|\Phi_{0}\rangle$. The energy unit is like in Eqn. (\ref{eq:zqham}). 
We have to minimize
\begin{eqnarray}
E(\langle {\mathcal T_{\bf
H}}\rangle)=\frac{1}{2}\left\langle{\mathcal T}_{\bf
H}\right\rangle^2+ E_{0}(\langle {\mathcal T}_{\bf H}\rangle)
\label{eq:en}
\end{eqnarray}
with respect to $\langle {\mathcal T}_{\bf H}\rangle$.   
In general, ${\mathcal T}_{\bf H}$ is not pointing in the
same direction in the $\Gamma_5$ space as the zero-field
${\mathcal T}=\langle{\mathcal T}_{111}^{\beta}\rangle$, but
neither is it collinear with ${\bf H}$; it has to be chosen in an
optimization procedure, observing the symmetry lowering due to
the magnetic field.

Fig.~\ref{fig:fields} shows results obtained by minimizing
$E({\mathcal T})$ with respect to ${\mathcal T}=\langle {\mathcal
T}_{111}^{\beta}\rangle$, our original choice of order parameter.
For ${\bf H}\parallel(111)$, the field does not introduce any
inequivalence of $x$, $y$, and $z$, thus the above choice of the
order parameter is optimal, and the second order transition seen
in Fig.~\ref{fig:fields} (left) is genuine.

Fig.~\ref{fig:fields} (right) shows the self-consistent solution
for ${\mathcal T}$ for fields ${\bf H}\parallel(123)$. (123) is taken
to represent general non-symmetric directions. We find behavior
characteristic of smeared-out phase transitions (the marked
upward curvature at $H\sim 1.5$ shows where the phase transition
might have been; clearly, intersite interactions are important
for $H<1.5$, while their effect is negligible in the high-field
tail).

\begin{figure}[ht] \centering
\includegraphics[totalheight=4.9cm,angle=270]{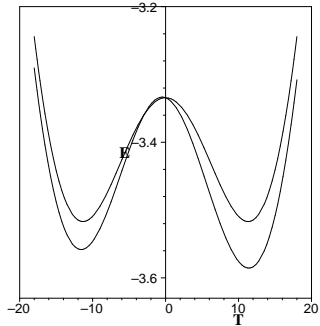}
\caption{The Landau-type ground state energy expression has
symmetric or asymmetric minima depending on whether the field is
applied in a symmetry direction (${\bf H}\parallel(111)$, upper curve),
or non-symmetry direction (${\bf H}\parallel(123)$, lower curve).}
\label{fig:fields2}
\end{figure}

The reason for the discrepancy between the ${\bf H}\parallel(111)$ and
${\bf H}\parallel(123)$ behavior becomes clear from plotting the
Landau-type ground state energy density for different field
directions (Fig.~\ref{fig:fields2}). For ${\bf H}\parallel(111)$,
equivalent minima remain at the positions $\pm{\cal T}_0$,
thus the system can pick one of these in a symmetry breaking
transition (upper curve). On the other hand, for
${\bf H}\parallel(123)$, the two minima are not equivalent, the ground
state remains always on the right-hand side. There is no symmetry
breaking transition though the $T$-dependence may be non-trivial,
showing the shadow of the phase transition which might have
happened.

One might object that for ${\bf H}\parallel(123)$, ${\mathcal T}_x$,
${\mathcal T}_y$, and ${\mathcal T}_z$ are no longer equivalent,
thus the optimal mean field solution should be sought in the form
\begin{equation}
{\mathcal T}_{\bf H} = r_x {\mathcal T}_x + r_y {\mathcal T}_y +
r_z {\mathcal T}_z\, . \label{eq:opttau}
\end{equation}
This is true but we do not have to make the considerable effort
of a three-parameter optimization. We will bring general
arguments to show that the solution would be like that in
Fig.~\ref{fig:fields} (right), whatever ${\mathcal T}_{\bf H}$ is
chosen. It remains true that for general (non-symmetric) field
directions, the ground state of ({\ref{eq:zham}) is non-degenerate
and therefore no symmetry breaking transition (in particular, no
continuous octupolar ordering transition) is possible.

For some symmetry directions which are not equivalent to (111)
(e.g., for (001)), the character of the solution is different
from either of those shown in Fig.~\ref{fig:fields}. We will
discuss these later.

\section{Symmetry analysis of field-induced multipoles}
\label{sec:symmetry}

\subsection{The high-field limit}
\label{sec:highfield}

We learn from Fig.~\ref{fig:fields} that at sufficiently high
fields the ground state is determined by the external field only:
either because ordering has been suppressed, or because there was
no transition to begin with. The following analysis of the
field-induced multipoles does not rely on the mean field
approximation, but each of the cases will be illustrated by a
mean field calculation.

The octupole operators are third-order polynomials of $J_x$, $J_y$
and $J_z$. The $\Gamma_5$ octupoles can be expressed in terms of
dipole and quadrupole operators \cite{sakai}
\begin{eqnarray}
{\cal T}_{x}^{\beta} &=& (\frac{1}{3}{\cal O}_{2}^{0}
+\frac{1}{6}{\cal O}_{2}^{2})J_{x}
+\frac{2}{3}({\cal O}_{zx}J_{z}-{\cal O}_{xy}J_{y})\nonumber \\
{\cal T}_{y}^{\beta} &=& (-\frac{1}{3}{\cal O}_{2}^{0}
+\frac{1}{6}{\cal O}_{2}^{2})J_{y}+\frac{2}{3}({\cal O}_{xy}J_{x}
-{\cal O}_{yz}J_{z})\nonumber \\
{\cal T}_{z}^{\beta} &=& -\frac{1}{3}{\cal O}_{2}^{2}J_{z}
+\frac{2}{3}({\cal O}_{yz}J_{y}-{\cal O}_{zx}J_{x})
\label{eq:tbetas}
\end{eqnarray}
where the quadrupoles are well-known quadratic expressions
\begin{eqnarray}
{\cal O}_2^0 & = & \frac{1}{2}\left( 2J_z^2-J_x^2-J_y^2\right)
 \nonumber \\
{\cal O}_2^2 & = & J_x^2-J_y^2 \nonumber\\
{\cal O}_{xy} & = & \frac{1}{2}\left(J_xJ_y + J_yJ_x\right) \nonumber  \\
{\cal O}_{yz} & = & \frac{1}{2}\left(J_yJ_z + J_zJ_y\right) \nonumber \\
{\cal O}_{zx} & = & \frac{1}{2}\left(J_zJ_x + J_xJ_z\right).
\end{eqnarray}

(\ref{eq:tbetas}) contains exact identities respecting the
non-commutative nature of the operators. However, ${\cal
T}_{z}^{\beta}$ etc. are themselves defined as symmetrized
expressions (\ref{eq:tbetadef}), so it must be true that the
order of the operators on the right-hand side cannot really
matter. In fact, there is an arbitrariness in the representation
(\ref{eq:tbetas}): it would be also true that
\begin{equation}
{\cal T}_{z}^{\beta} =  -\frac{1}{3}{\overline {{\cal O}_{2}^{2}J_{z}}}
\label{eq:tbeta1}
\end{equation}
or
\begin{equation}
{\cal T}_{z}^{\beta} =  \frac{2}{3}\left({\overline {{\cal O}_{yz}J_{y}}} -
{\overline {{\cal O}_{zx}J_{x}}}\right)\,.
\label{eq:tbeta2}
\end{equation}
Similar relationships can be listed for the first two lines of
(\ref{eq:tbetas}). This suggests that the relationships
(\ref{eq:tbetas}) can also be interpreted in terms of
$c$-numbers, i.e., classical polarization densities \cite{tmte}.
Such considerations are valid for either field-induced, or
interaction-induced multipole densities. This enables us to use
relationships like (\ref{eq:tbetas}) in Landau expansions.

First, we discuss field-induced densities. The basic idea is
that an external field induces $\langle{\bf J}\rangle
=(\langle J_x\rangle,\langle J_y\rangle,\langle J_z\rangle)\parallel{\bf H}$ ,
and this gives rise to induced quadrupoles $\langle{\cal O}_{xy}\rangle=
\langle J_x\rangle\langle J_y\rangle$ as a second-order effect, and induced
octupoles as a third-order effect, etc. If an octupole component is
field-induced, it can no longer play the role of the order parameter of a
symmetry breaking transition. The question is, can it happen that certain
octupole moments are {\sl not} induced by the field.

(\ref{eq:tbeta1}) and (\ref{eq:tbeta2}) are still separate operator
identities, but they must have the same classical meaning when $J_x$, etc.
are treated as $c$-numbers. Indeed from (\ref{eq:tbeta2})
\begin{eqnarray}
{\cal T}_{z}^{\beta} & = & {\overline {{\cal O}_{yz}J_{y}}}
-{\overline {{\cal O}_{zx}J_{x}}}
\longrightarrow (\langle J_y\rangle\langle J_z\rangle)\langle J_y\rangle -
(\langle J_z\rangle\langle J_x\rangle)\langle J_x\rangle\nonumber\\
& = & (\langle J_y\rangle^2 -\langle J_x\rangle^2 )\langle J_z\rangle
\propto (H_y^2 -H_x^2 )H_z
\label{eq:amk1}
\end{eqnarray}
which is the same that we would have obtained from (\ref{eq:tbeta1}).
Similarly,
\begin{eqnarray}
{\cal T}_{x}^{\beta} & = & {\overline {{\cal O}_{zx}J_{z}}} -
{\overline {{\cal O}_{xy}J_{y}}}
\longrightarrow (\langle J_z\rangle\langle J_x\rangle)\langle J_z\rangle -
(\langle J_x\rangle\langle J_y\rangle)\langle J_y\rangle\nonumber\\
& = & (\langle J_z\rangle^2 -\langle J_y\rangle^2 )\langle J_x\rangle
\propto (H_z^2 - H_y^2 ) H_x \, ,
\label{eq:amk2}
\end{eqnarray}
and
\begin{eqnarray}
{\cal T}_{y}^{\beta} & = & {\overline {{\cal O}_{xy}J_{x}}}
-{\overline {{\cal O}_{yz}J_{z}}}
\propto ( H_x^2 - H_z^2 ) H_y\, .
\label{eq:amk3}
\end{eqnarray}

 We can also argue in the following manner. Higher-order
polarizations in a magnetic field give rise to the following
general $H$-dependence of the energy
\begin{equation}
{\cal E}(H) \sim {\cal E}(H=0) - \frac{\chi}{2} H^2 -
\frac{\chi_3}{12} H^4 \,...
\end{equation}
The lowest order time reversal invariant expression containing 
${\cal T}$ is ${\cal T}H$, thus the coupling of octupolar
moments to fields may appear in terms from $O(H^4)$
upwards. If it does, the minimal eigenvalue of the mean field
energy (\ref{eq:en0}) will not be symmetrical under the sign
change of octupole moments: $E_0({\cal T})\ne E_0(-{\cal T})$.
Non-equivalent minima like in Fig.~\ref{fig:fields2} (lower curve) 
mean that there is no symmetry to break, a phase
transition is not possible. However, for fields in the special
directions discussed above, there is no field-induced $\Gamma_5$
octupole \cite{footnote}, the $\pm{\cal T}$ minima of
$E_0({\cal T})$ remain equivalent (Fig.~\ref{fig:fields2},
upper curve) and spontaneous symmetry breaking remains possible.
The eventual merging of the two minima is no longer a question of
symmetry, but of field intensity; a sufficiently strong field will
suppress octupolar (or any other) order, and produce a unique
polarized state for any field direction (Fig.~\ref{fig:fields}, 
right).

In what follows, we calculate the induced $\Gamma_5$ octupoles
using (\ref{eq:amk1})--(\ref{eq:amk3}) for several field
directions, and discuss the possibility of symmetry breaking
transitions.

\subsubsection{Non-symmetric directions}

First let us observe that a field pointing in a general direction
will give non-zero values for ${\cal T}_{x}^{\beta}$, 
${\cal T}_{y}^{\beta}$, and ${\cal T}_{z}^{\beta}$. 
Since the field induces all $\Gamma_5$
octupoles, there remains no degeneracy to be lifted, no symmetry
breaking transition is possible \cite{Ising}. This corresponds to
the situation in the right-hand panel of Fig.~\ref{fig:fields}.

\subsubsection{$H\parallel(111)$}

Taking now ${\bf H}\parallel(111)$, we find
${\cal T}_{x}^{\beta}={\cal T}_{y}^{\beta}={\cal T}_{z}^{\beta}=0$, and so also
${\cal T}_{111}^{\beta}=0$. The field does not induce $\Gamma_5$
octupoles, and therefore a symmetry breaking transition is
possible. Furthermore, as we remarked earlier, the $x$, $y$, and
$z$ axes play equivalent roles, and therefore the choice of the
order parameter ${\mathcal T}=\langle{\mathcal
T}_{111}^{\beta}\rangle$ is correct. The situation corresponds to
Fig.~\ref{fig:fields} (left).

\subsubsection{$H\parallel(001)$}

Next consider ${\bf H}\parallel(001)$. Also here we find
${\cal T}_{x}^{\beta}={\cal T}_{y}^{\beta}={\cal T}_{z}^{\beta}=0$ from Eqs.
(\ref{eq:amk1})--(\ref{eq:amk3}), and therefore the possibility of
continuous phase transitions. However, it is intuitively clear
that the $z$-axis is inequivalent to $x$ and $y$, and therefore
the order parameter may be either ${\mathcal T}_z$, or some
linear combination of ${\mathcal T}_x$ and ${\mathcal T}_y$. We
have to perform a two-parameter minimization using the suitably
modified form of ({\ref{eq:en})
\begin{equation}
E_0(\langle {\mathcal T}_z\rangle,\langle {\mathcal T}_{\perp}
\rangle) = -J \left( ({\mathcal T}_x+{\mathcal T}_y)\langle
{\mathcal T}_{\perp}\rangle +  {\mathcal T}_z\langle{\mathcal
T}_z\rangle \right)\label{eq:en001}
\end{equation}
with $\langle {\mathcal T}_x\rangle=\langle {\mathcal
T}_y\rangle=\langle {\mathcal T}_{\perp}\rangle$. Like in
Fig.~\ref{fig:fields2}, we expect that the ground state energy
functional has degenerate local minima: at low fields, we find a
pair of these as a function of $\langle {\mathcal T}_z\rangle$,
and another pair along the $\langle {\mathcal T}_x+{\mathcal
T}_y\rangle$ direction (the latter choice is arbitrary in the
sense that we could also have taken $\langle {\mathcal
T}_x-{\mathcal T}_y\rangle$) (Fig.~\ref{fig:contour_100}, top
left). These two pairs of minima are not symmetry-related, as it
is also shown by the fact that at intermediate fields, only the
$\langle {\mathcal T}_x+{\mathcal T}_y\rangle\ne 0$ minima survive
(Fig.~\ref{fig:contour_100}, top right). At high fields, the
ground state is non-degenerate with $\langle {\mathcal
T}_x+{\mathcal T}_y\rangle=\langle {\mathcal T}_z\rangle=0$
(Fig.~\ref{fig:contour_100}, bottom).

\begin{figure}
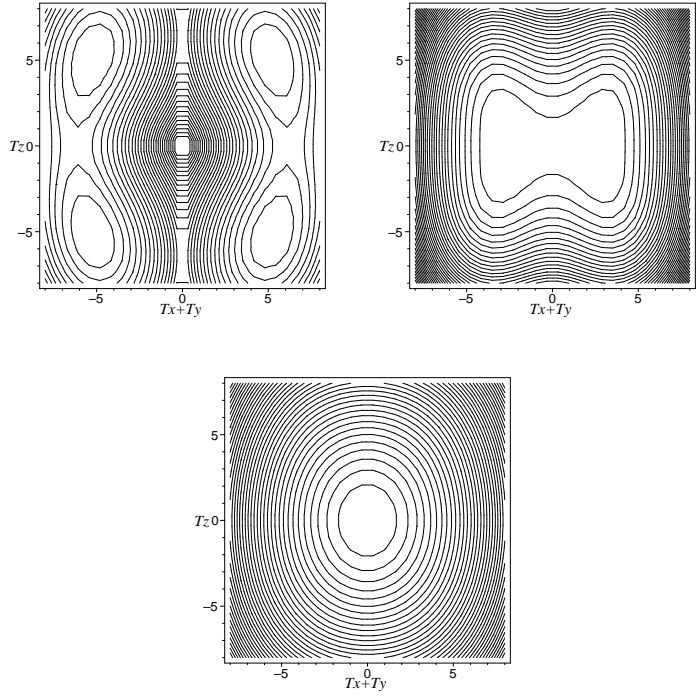

\centering
\includegraphics[totalheight=4.8cm,angle=270]{akpf_fig_8_topleft.ps}
\includegraphics[totalheight=4.8cm,angle=270]{akpf_fig_8_topright.ps}
\includegraphics[totalheight=4.8cm,angle=270]{akpf_fig_8_bottom.ps}
\caption{The contour plot of the ground state energy functional in
the
$\left<{\cal T}_{x}^{\beta}+{\cal T}_{y}^{\beta}\right>$--
$\left<{\cal T}_{z}\right>$
plane for ${\bf H}=(0,0,H)$ magnetic fields $H=0.2$ ({\sl top
left}), $H=0.42$ ({\sl top right}) and $H=0.8$ ({\sl
bottom}).}\label{fig:contour_100}
\end{figure}

The corresponding sequence of two second-order ground state
transitions is shown in Fig.~\ref{fig:H100} (left). $\langle
{\cal T}_{x}^{\beta}+{\cal T}_{y}^{\beta}\rangle >0$ (or alternatively, 
$\langle {\cal T}_{x}^{\beta}-{\cal T}_{y}^{\beta}\rangle >0$) 
order develops at higher
critical field $H_{\rm cr}^>$,  with 
$\langle {\cal T}_{z}^{\beta}\rangle =0$. Upon reducing the field to a lower
critical value $H_{\rm cr}^<$, a second symmetry breaking occurs.
In the low-field phase $H<H_{\rm cr}^<$, $\langle {\cal T}_{x}^{\beta}
+{\cal T}_{y}^{\beta}\rangle\ne 0$ and also $\langle {\cal T}_{z}^{\beta}
\rangle \ne 0$. $ \langle {\cal T}_{z}^{\beta}\rangle\ne
\langle {\cal T}_{x}^{\beta}+{\cal T}_{y}^{\beta}\rangle/2$ as long as $H>0$;
the ${\mathcal T}_{111}$ order ($ \langle
{\cal T}_{x}^{\beta}\rangle=\langle {\cal T}_{y}^{\beta}\rangle=\langle
{\cal T}_{z}^{\beta}\rangle$)  appears continuously as $H\to 0$.

\begin{figure}
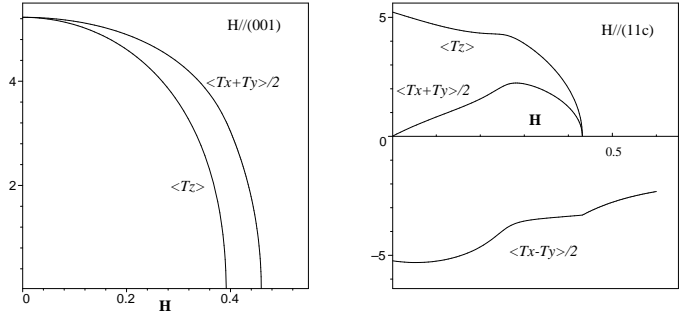

\centering
\includegraphics[totalheight=4.8cm,angle=270]{akpf_fig_9_left.ps}
\includegraphics[totalheight=4.8cm,angle=270]{akpf_fig_9_right.ps}
\caption{{\sl Left}: The field dependence of $\langle
{\cal T}_{x}^{\beta}+{\cal T}_{y}^{\beta}\rangle/2$ and 
$\langle {\cal T}_{z}\rangle$
in a field ${\bf H}\parallel(001)$. {\sl Right}: Octupolar components for 
${\bf H}\parallel(11c)$, $c\ne 0$.} \label{fig:H100}
\end{figure}

\subsubsection{$H\parallel(11c)$}

${\bf H}\parallel(11c)$ ($c\ne 0$) induces $\langle
{\cal T}_{x}^{\beta}-{\cal T}_{y}^{\beta}\rangle\ne 0$, leaving
$\left<{\cal T}_{x}^{\beta}+{\cal T}_{y}^{\beta}\right>=0$ and
 $\langle {\cal T}_{z}^{\beta}\rangle=0$. 
There is a remaining octupolar degeneracy
which is is lifted in a single continuous transition, where 
$\langle{\mathcal T}_{z}^{\beta}\rangle\ne 0$, and 
$\langle{\mathcal T}_{x}^{\beta}+{\mathcal T}_{y}^{\beta}\rangle\ne 0$ 
appear simultaneously
(Fig.~\ref{fig:H100}, right). For $c=0$, (i.e., ${\bf H}\parallel(110)$) 
${\mathcal T}_{z}^{\beta}$, and 
${\mathcal T}_{x}^{\beta}+{\mathcal T}_{y}^{\beta}$ can order independently,
like in the case of ${\bf H}\parallel(001)$. 

\subsubsection{Field direction dependence: Summary}

We have discussed field directions which do not subtend a too
large angle with (111), thus it holds that the limit $H\to 0$
picks the (111) ground state. For other field directions, the
limit $H\to 0$ may give one of the other ground states, e.g.
(1${\overline 1}$1) type order (see (\ref{eq:111states})).

The previously discussed special directions which allowed a
symmetry breaking transition, were all lying in the plane with
normal vector ${\overrightarrow n}=(1,-1,0)$. Because of the cubic
symmetry, the behaviour is the same for magnetic fields lying in
planes with normal vectors ${\overrightarrow n}=(1,1,0)$,
${\overrightarrow n}=(1,0,-1)$, ${\overrightarrow
n}=(1,0,1)$,${\overrightarrow n}=(0,1,-1)$, and ${\overrightarrow
n}=(0,1,1)$, only the ordering phases change correspondingly.
These six planes intersect along the directions
$\left[111\right]$, $\left[{\overline 1}11\right]$,
$\left[1{\overline 1}1\right]$ and $\left[11{\overline
1}\right]$ (Fig.~\ref{fig:planes}). Any direction outside 
these planes excludes the
possibility of a continuous octupolar transition.

\begin{figure}
\centering
\includegraphics[totalheight=5cm,angle=270]{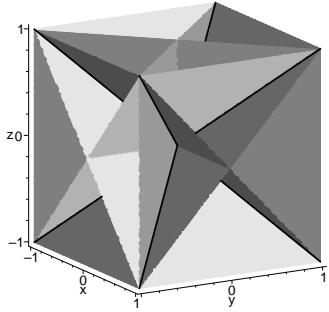}
\caption{Field directions lying in any of the planes shown allow a continuous
  octupolar ordering transition. Special rules hold for the lines of
  intersection, and other high-symmetry directions.} 
\label{fig:planes}
\end{figure}

\subsection{Field-induced multipoles}

We may also regard the problem of field-induced $\Gamma_5$ octupoles as a
special case of field-induced multipoles in general
\cite{shiina,shiina98}. It is best to begin with quadrupoles.
When the magnetic field is zero, the $\Gamma_{5}$-type
quadrupolar moments ${\cal O}_{xy}$, ${\cal O}_{zx}$, ${\cal
O}_{yz}$ are equivalent by cubic symmetry. Switching on an
external magnetic field with general direction
$\overrightarrow{H}\|\overrightarrow{n}$, where
$\overrightarrow{n}= (\kappa,\lambda,\mu)$, the quadrupolar
operator along the $\overrightarrow{n}$ direction can be given as
\begin{eqnarray}
 Q(\overrightarrow{n}) &\approx&
 3(\overrightarrow{n}\overrightarrow{J})^2-J(J+1)=
3\kappa\lambda(J_{x}J_{y}+J_{y}J_{x})+3\kappa\mu(J_{x}J_{z}+J_{z}J_{x})
 \nonumber \\
& &  +3\lambda\mu(J_{x}J_{y}+J_{y}J_{x})+
\kappa^2J_{x}^2+\lambda^2J_{y}^2+\mu^2J_{z}^2)-J(J+1) \nonumber \\
& = & 3\kappa\lambda{\mathcal O}_{xy}+3\kappa\mu{\mathcal O}_{zx}
+3\lambda\mu{\mathcal O}_{yz}+
(\mu^2-\lambda^2)(J_{z}^2-J_{y}^2) \nonumber \\
& & +(\kappa^2-\mu^2)(J_{x}^2-J_{z}^2)+(\lambda^2-\kappa^2)(J_{y}^2-J_{x}^2)
 \nonumber \\
& & +(\kappa^2+\lambda^2+\mu^2-1)J(J+1)\, . \label{eq:genquad}
\end{eqnarray}
This means that the ${\cal O}_{xy}$, ${\cal O}_{zx}$ and ${\cal O}_{yz}$
$\Gamma_{5}$-type quadrupolar operators are no longer equivalent,
the quadrupolar moment is distorted along the external magnetic
field direction. Above, we obtained the components of the quadrupolar operator
in the rotated new basis. The $\Gamma_{5}$
quadrupole moments ${\mathcal O}_{xy}$, ${\mathcal O}_{zx}$, ${\mathcal
  O}_{yz}$ are proportional to $\kappa\lambda\sim H_{x}H_{y}$,
$\kappa\mu\sim H_{x}H_{z}$, $\lambda\mu\sim H_{y}H_{z}$. The next three
quadrupolar terms, which are linear combinations of the two
$\Gamma_{3}$ quadrupoles, are shown because they appear in the expressions of
the ${\cal T}^{\beta}$ octupoles:
$J_{z}^2-J_{y}^2\sim H_{z}^2-H_{y}^2$, $J_{x}^2-J_{z}^2\sim H_{x}^2-H_{z}^2$,
$J_{y}^2-J_{x}^2\sim H_{y}^2-H_{x}^2$. The last term is an invariant number.

\subsection{Ordering in external magnetic field}
\label{sec:helmholtz}

Now we consider arbitrary field intensities. At sufficiently low fields,
symmetry breaking transitions are possible. However, the magnetic field 
lowers the symmetry of the system in a peculiar way, and gives rise to 
couplings between order parameters which would be independent in the absence
of a field. The nature of these couplings depends sensitively on field
direction.

Our previous discussion was about the ground state energy $E_0(y,{\bf H})$,
where {\bf H} is the external magnetic field, and $y$ stands for all other
variables. The magnetic moment, or in our case ${\bf J}$, is obtained as 
${\bf J}=-(\partial E_0/\partial {\bf H})_y$. 

In what follows, we prefer to use the Helmholtz free energy  ${\cal G}$ which
is related to $E_0$ by the Legendre transformation ${\cal G}=E_0+{\bf
  J}{\cdot}{\bf H}$. The magnetic field will be expressed as 
\begin{equation}
{\bf H} = \frac{\partial {\cal G}}{\partial {\bf J}} \, .
\label{eq:helmH}
\end{equation}

The generalized Helmholtz free energy can be expanded in terms of
the components of the symmetry-allowed multipoles \cite{shiina}
\begin{eqnarray}
{\cal G} & = & {\cal
G}(J_x,J_y,J_z;{\cal O}_2^2,{\cal O}_2^0,{\cal O}_{xy},{\cal O}_{yz},
{\cal O}_{zx};{\cal
T}_{xyz},{\cal T}_{x}^{\alpha},{\cal T}_{y}^{\alpha},{\cal
T}_{z}^{\alpha},{\cal T}_{x}^{\beta},{\cal T}_{y}^{\beta},{\cal
T}_{z}^{\beta})\nonumber\\
& = & \sum_{i,j,...} {\cal I}(\Gamma_i{\otimes}\Gamma_j...) 
\label{eq:helm1}
\end{eqnarray}
We have to go over the list of all possible product
representations spanned by the order parameter components, 
and identify the bases for the identity representation 
$\Gamma_1$; these are the invariants 
${\cal I}(\Gamma_i{\otimes}\Gamma_j...)$ (assuming that 
they are also time-reversal invariant).

It is obvious that the present argument is valid only for phases with uniform
order. ${\bf q}\ne 0$ Fourier components of the
multipole densities should be included in a Landau theory of modulated states,
such as the antiferro-octupolar phase observed in NpO$_2$.

Returning to (\ref{eq:helm1}): to construct a Landau theory, 
we would need all the invariants up to some specified
order, but to derive {\bf H}, it is enough to consider those 
which contain the components of {\bf J}. 
Their general form is ${\cal I}(\Gamma_4{\otimes}\Gamma_j...)=
{\bf J}{\cdot}{\bf V}$, where the components of {\bf V} give the basis of 
$\Gamma_{4u}$ ($g$ and $u$ refer to parity under time reversal). 

We arrange the invariants according to the number of factors in the underlying
product representation. For the present purposes, we will call this number the
order of the invariant \cite{order}.  
The second order invariants containing {\bf J} 
are ${\bf J}{\cdot}{\bf J}$ and ${\bf
J}{\cdot}{\vec {\cal T}}^{\alpha}$. Third order invariants arise from
$\Gamma_4{\otimes}\Gamma_5{\otimes}\Gamma_5$,
$\Gamma_4{\otimes}\Gamma_3{\otimes}\Gamma_5$,
$\Gamma_4{\otimes}\Gamma_2{\otimes}\Gamma_5$, and
$\Gamma_4{\otimes}\Gamma_4{\otimes}\Gamma_5$. To take the
simplest example, consider
$\Gamma_4{\otimes}\Gamma_2{\otimes}\Gamma_5$, for which
\begin{equation}
{\cal I}(\Gamma_4{\otimes}\Gamma_2{\otimes}\Gamma_5) = 
J_x{\cal O}_{yz}{\cal T}_{xyz}+J_y {\cal O}_{zx}{\cal T}_{xyz}+
J_z {\cal O}_{xy}{\cal T}_{xyz}\, ,
\label{eq:helm2}
\end{equation}
and the corresponding term of {\bf V} is
\begin{equation}
{\bf
V}(\Gamma_2{\otimes}\Gamma_5)=(O_{yz}{\cal T}_{xyz},{\cal O}_{zx}
{\cal T}_{xyz},{\cal O}_{xy}{\cal T}_{xyz})\, .
\label{eq:helm3}
\end{equation}
Taking e.g., the $z$-component, we find that the magnetic field couples to
${\cal O}_{xy}{\cal T}_{xyz}$. One possible interpretation is that, in the
presence of ${\bf H}\parallel(001)$, ${\cal O}_{xy}$-type quadrupolar moment
induces the octupole ${\cal T}_{xyz}$ \cite{shiina}. 
Alternatively, ${\cal T}_{xyz}$-type
octupole order would induce  ${\cal O}_{xy}$ quadrupoles. 

In our further discussion of third order invariants, we confine our attention
to those which have a bearing on the question of $\Gamma_5$ octupolar order,
i.e., one of the factors is $\Gamma_{5u}$. As for 
$\Gamma_4{\otimes}\Gamma_4{\otimes}\Gamma_5$ invariants, since one of the 
$\Gamma_4$ has to give {\bf J}, i.e., it is $\Gamma_{4u}$, the remaining 
$\Gamma_4$ must be $\Gamma_{4g}$. The lowest order $\Gamma_{4g}$ multipole is
a hexadecapole. However, within our $\Gamma_8$ subspace, hexadecapoles cannot 
be independent of the first 15 multipoles, thus the formally third order
expression would have to be rewritten as a fourth-order invariant. We
generally neglect terms of fourth order, and seek to draw conclusions from the
genuinely third order terms. These belong to 
$\Gamma_{4u}{\otimes}\Gamma_{5u}{\otimes}\Gamma_{5g}$ which gives 
\begin{eqnarray} 
{\cal I}(\Gamma_{4u}{\otimes}\Gamma_{5u}{\otimes}\Gamma_{5g}) & = &
J_x(-{\cal O}_{xy}{\cal T}_y^{\beta}+{\cal O}_{zx}{\cal T}_z^{\beta}) + 
J_y(-{\cal O}_{yz}{\cal T}_z^{\beta}+{\cal O}_{xy}{\cal T}_x^{\beta})
\nonumber\\
& & + 
J_z(-{\cal O}_{zx}{\cal T}_x^{\beta}+{\cal O}_{yz}{\cal T}_y^{\beta})\, ,
\label{eq:helm4}
\end{eqnarray}
and to 
$\Gamma_{4u}{\otimes}\Gamma_{5u}{\otimes}\Gamma_{3g}$ which gives 
\begin{eqnarray}
{\cal I}(\Gamma_{4u}{\otimes}\Gamma_{5u}{\otimes}\Gamma_{3g}) = 
-\frac{1}{2} J_x ({\cal O}_2^0 +{\cal O}_2^2)
{\cal T}_x^{\beta}  
+\frac{1}{2} J_y ({\cal O}_2^0 -{\cal O}_2^2)
{\cal T}_y^{\beta} +  J_z{\cal O}_2^2{\cal T}_z^{\beta}\, .
\label{eq:helm5}
\end{eqnarray}
The invariants (\ref{eq:helm4}) and (\ref{eq:helm5}) appear with the 
independent coefficients $w_1$ and $w_2$ in ${\cal G}$. For the $z$-component
of the field we give first a fuller expression derived from a number of 
low-order invariants 
\begin{eqnarray}
H_z & = & u_1 J_z + u_2 {\cal T}_z^{\alpha} + 
v_1 J_z {\cal O}_2^0 + v_2 {\cal O}_{xy}{\cal T}_{xyz}\nonumber\\
& &  + 
w_1(-{\cal O}_{zx}{\cal T}_x^{\beta}+{\cal O}_{yz}{\cal T}_y^{\beta})
+w_2{\cal O}_2^2{\cal T}_z^{\beta} + ...  
\label{eq:helm6a}
\end{eqnarray}
Terms in the first line are needed to recover the results by \cite{shiina}. 
However, we are now only interested in the interplay of $\Gamma_5$ octupoles 
and fields, therefore we omit from (\ref{eq:helm6a}) terms not containing 
${\cal T}_i^{\beta}$
\begin{eqnarray}
H_z = w_1(-{\cal O}_{zx}{\cal T}_x^{\beta}+{\cal O}_{yz}{\cal T}_y^{\beta})
+w_2{\cal O}_2^2{\cal T}_z^{\beta} \, .  
\label{eq:helm6}
\end{eqnarray}
If quadrupolar interactions induce any of the quadrupolar moments appearing in
the above equation, the field will induce $\Gamma_5$ octupoles, thus
explicitely breaks the symmetry of the problem. However, in the absence of
such interactions, we can turn to the high-field expressions 
(\ref{eq:amk1})--(\ref{eq:amk3}) which give none of these
quadrupoles. Therefore, symmetry breaking by octupolar ordering is a
possibility.

For fields of other orientation, we need also the following relationships
\begin{eqnarray}
H_x = w_1(-{\cal O}_{xy}{\cal T}_y^{\beta}+{\cal O}_{zx}{\cal T}_z^{\beta})
-\frac{w_2}{2} ({\cal O}_2^0 +{\cal O}_2^2)
{\cal T}_x^{\beta} \, ,
\label{eq:helm7}
\end{eqnarray}
and
\begin{eqnarray}
H_y = w_1(-{\cal O}_{yz}{\cal T}_z^{\beta}+{\cal O}_{xy}{\cal T}_x^{\beta})
+\frac{w_2}{2} ({\cal O}_2^0 -{\cal O}_2^2)
{\cal T}_y^{\beta} \, .
\label{eq:helm8}
\end{eqnarray}

For a field in the (111) direction, 
\begin{eqnarray}
H_{111} & = & w_1\left[ \left ({\cal O}_{xy} -{\cal O}_{zx}\right){\cal T}^{\beta}_x + 
\left({\cal O}_{yz} -{\cal O}_{xy}\right){\cal T}^{\beta}_y + 
\left({\cal O}_{zx} -{\cal O}_{yz}\right){\cal T}^{\beta}_z \right]\\
& & + w_2\left[ -\frac{1}{2} \left({\cal O}_2^0 +{\cal O}_2^2\right)
{\cal T}_x^{\beta}  
+\frac{1}{2} \left({\cal O}_2^0 -{\cal O}_2^2\right)
{\cal T}_y^{\beta} +  {\cal O}_2^2{\cal T}_z^{\beta}\right]\, . \nonumber
\end{eqnarray}
Neither of the quadrupolar coefficients seen above are field-induced. 
Therefore, if there is no quadrupolar interaction to introduce some of 
them as order parameters, ${\bf H}\parallel(111)$ fields will allow the 
same kind of $\Gamma_5$ octupolar ordering as in the absence of a field 
(remember, though, that the amplitude of the order will be gradually 
suppressed by the field). 
 
In a (110) field
\begin{eqnarray}
H_{110} & = &  -\frac{w_2}{2} {\cal O}_2^0 
\left({\cal T}_x^{\beta} -{\cal T}_y^{\beta}\right) \nonumber\\[2mm]
& & - \frac{w_2}{2} {\cal O}_2^2 \left({\cal T}_x^{\beta} 
+{\cal T}_y^{\beta}\right)
+ w_1 \left( {\cal O}_{zx} -{\cal O}_{yz} \right){\cal T}_z^{\beta}\, .
\end{eqnarray}
The point to note from (\ref{eq:amk1})--(\ref{eq:amk3}) is that though 
$({\cal O}_{zx} -{\cal O}_{yz})$ and ${\cal O}_2^2$ are not induced 
by the field, 
${\cal O}_{2}^0$ is, and therefore the octupolar component 
$({\cal T}_x^{\beta} -{\cal T}_y^{\beta})$ is also field induced. The remaining 
octupolar degeneracy arises from the fact that   
$({\cal T}_x^{\beta} +{\cal T}_y^{\beta})$ and ${\cal T}_z^{\beta}$ do not 
couple to the field. Since these are associated with different terms in the expansion 
(\ref{eq:helm1}), $({\cal T}_x^{\beta} +{\cal T}_y^{\beta})$ and ${\cal T}_z^{\beta}$ 
may order independently.

Finally, we comment upon the case ${\bf H}\parallel(11c)$ 
\begin{eqnarray}
H_{11c} & = & w_1\left[ \left ({\cal O}_{xy} -
c{\cal O}_{zx}\right){\cal T}^{\beta}_x + 
\left(c{\cal O}_{yz} -{\cal O}_{xy}\right){\cal T}^{\beta}_y + 
\left({\cal O}_{zx} -{\cal O}_{yz}\right){\cal T}^{\beta}_z \right]
\nonumber\\
& & + w_2\left[ -\frac{1}{2} \left({\cal O}_2^0 +{\cal O}_2^2\right)
{\cal T}_x^{\beta}  
+\frac{1}{2} \left({\cal O}_2^0 -{\cal O}_2^2\right)
{\cal T}_y^{\beta} +  c{\cal O}_2^2{\cal T}_z^{\beta}\right]\, .
\end{eqnarray}
Again only $({\cal T}_x^{\beta} -{\cal T}_y^{\beta})$ is field induced. However, once 
octupole--octupole interaction gives rise to ${\cal T}_z^{\beta}$ order, it induces 
${\cal O}_2^2$, which in turn induces $({\cal T}_x^{\beta} +{\cal T}_y^{\beta})$, thus 
there is a single phase transition (Fig.~\ref{fig:H100}, right).

\section{Conclusion}

The $\Gamma_8$ subspace supports a variety of competing order parameters. The
fourfold ground state degeneracy can be lifted either in two steps (removing
Kramers and non-Kramers degeneracies separately), or in a single phase
transition. The latter possibility is realized by the ordering of $\Gamma_5$
octupoles. We have found that the crystal field gives rise to a peculiar
single-ion octupolar anisotropy which makes the choice of
${\mathcal T}_{111}^{\beta}= {\mathcal T}_x^{\beta}+{\mathcal T}_y^{\beta}
+{\mathcal T}_z^{\beta}$ octupoles preferable as order parameter.  Though it breaks time reversal invariance,
octupolar ordering is non-magnetic in the sense of yielding vanishing dipole
moments. On the other hand, the ordering of $\Gamma_5$ octupoles induces
$\Gamma_5$ quadrupoles as secondary order parameter; this feature allows the
simultaneous lifting of Kramers and non-Kramers degeneracies.

Our discussion is mainly about a hypothetical $\Gamma_5$-type ferro-octupolar
ordering in a lattice of $\Gamma_8$ shells. The mean-field results presented
in Fig.~\ref{fig:oqphased} and Fig.~\ref{fig:oqop} are equally valid for 
ferro-octupolar, and the triple-${\vec k}$ antiferro-octupolar, ordering
patterns, but our main interest is in magnetic field effects, and our
arguments for the case of non-zero magnetic field are restricted to 
uniform phases. 
 
Magnetic octupoles are not time reversal invariant, thus we might have 
expected that spontaneous symmetry breaking due to uniform octupolar 
ordering is necessarily suppressed by magnetic fields. 
Indeed, for fields of a general orientation, we find that the
degeneracy of different octupolar ground states is immediately
lifted, and the para-octupolar state prevails at all
temperatures. However, the analysis of field-induced multipoles
shows that for field directions lying in certain planes, the
field does not induce all the ${\mathcal T}^{\beta}$-type
octupolar moments, and therefore sharp octupolar transitions
remain possible up to a certain critical field. The size of the
critical field, and the nature of the transition to the
high-field dipolar state, depend on the details of field
orientation within the symmetry-specified planes.

{\bf Acknowledgements}. The authors are greatly indebted to Prof. Hiroyuki 
Shiba for enlightening discussions and correspondence.  Shinsaku Kambe 
awoke our interest in the problem of NpO$_2$. At some points, we used a 
program developed by Katalin Radn\'oczi, and enjoyed discussions about various 
aspects of the problem with Karlo Penc. We were 
supported by the Hungarian national grants OTKA T 038162, OTKA T 037451, 
and OTKA TS 040878. A.K. is supported by the Budapest--Marburg European 
Graduate College {\sl Electron-Electron Interactions in Solids}. Both authors 
 enjoyed hospitality at 
EPFL Lausanne, and the Philipps Universit\"at Marburg during several 
short stays while this work was in progress.  
\\[5mm]

\appendix
\section{Numerical coefficients of cubic crystal field levels in the $\Gamma_{8}$ basis\label{appendix:a}}

\begin{eqnarray}
\alpha &= &\frac{\frac{26}{3}+\frac{1}{6}\sqrt{206}\sqrt{14}}{\sqrt{1+\left(\frac{26}{3}+
\frac{1}{6}\sqrt{206}\sqrt{14}\right)^2+\left(-\frac{5}{6}\sqrt{14}-\frac{1}{6}\sqrt{206}\right)^2}}=0.9530\nonumber \\
\beta &= &\frac{-\frac{5}{6}\sqrt{14}-\frac{1}{6}\sqrt{206}}{\sqrt{1+\left(\frac{26}{3}+
\frac{1}{6}\sqrt{206}\sqrt{14}\right)^2+\left(-\frac{5}{6}\sqrt{14}-\frac{1}{6}\sqrt{206}\right)^2}}=-0.2980\nonumber \\
\gamma &= &\frac{1}{\sqrt{1+\left(\frac{26}{3}+
\frac{1}{6}\sqrt{206}\sqrt{14}\right)^2+\left(-\frac{5}{6}\sqrt{14}-\frac{1}{6}\sqrt{206}\right)^2}}=0.05409\nonumber \\
\delta &= &\frac{-\frac{1}{15}\sqrt{14}\sqrt{6}-\frac{1}{30}\sqrt{6}\sqrt{206}}{\sqrt{1+\left(
-\frac{1}{15}\sqrt{14}\sqrt{6}-\frac{1}{30}\sqrt{6}\sqrt{206}\right)^2}}=-0.8721\nonumber \\
\epsilon &= &\frac{1}{\sqrt{1+\left(
-\frac{1}{15}\sqrt{14}\sqrt{6}-\frac{1}{30}\sqrt{6}\sqrt{206}\right)^2}}=0.4891\nonumber
\end{eqnarray}

The Zeeman splitting parameters appearing in Sec. \ref{sec:susc} are
\begin{eqnarray}
g_H & = & \frac{3}{206} \sqrt{ 129471 + 618\sqrt{206}\sqrt{14}} = 5.736
\nonumber\\[2mm]
y_H & = & \frac{1}{618} \sqrt{ 1078719 - 6798\sqrt{206}\sqrt{14}} = 1.3665
\nonumber
\end{eqnarray}

\section{Multipole moments in the ground state of ${\mathcal T}_{111}$
\label{appendix:b}}

\begin{eqnarray}
A & = &\left\langle\phi_{0}\left|{\mathcal T}_{111}\right|\phi_{0}\right\rangle=
\left\langle\phi_{0}\left|{\mathcal T}_{{\overline 1}11}\right|\phi_{0}\right\rangle=
\left\langle\phi_{0}\left|{\mathcal T}_{1{\overline 1}1}\right|\phi_{0}\right\rangle=
\left\langle\phi_{0}\left|{\mathcal T}_{11{\overline 1}}\right|\phi_{0}\right\rangle= \nonumber \\
& =&-\frac{15}{103}\sqrt{22660-206\sqrt{206}\sqrt{14}}=-15.683\nonumber \\
B/3 & = & \left\langle\phi_{0}\left|{\cal O}_{xy}\right|\phi_{0}\right\rangle
=\left\langle\phi_{0}\left|{\cal O}_{yz}\right|\phi_{0}\right\rangle=
\left\langle\phi_{0}\left|{\cal O}_{xz}\right|\phi_{0}\right\rangle=2.673
\nonumber\\
m&=&\left\langle\phi_{0}\left|J_{x}\right|\phi_{0}\right\rangle=
\left\langle\phi_{0}\left|J_{y}\right|\phi_{0}\right\rangle=
\left\langle\phi_{0}\left|J_{z}\right|\phi_{0}\right\rangle=0\nonumber
\end{eqnarray}


\begin{thebibliography}{99}
\bibitem{gamma5} Though as far as the subgroup of proper rotations is 
concerned, the symmetry of $\Gamma_5$ quadrupoles and 
$\Gamma_5$ octupoles is the same, they have different parity under 
time reversal. Their coupling is made possible by the invariant arising 
from $\Gamma_{5u}{\otimes}\Gamma_{5u}{\otimes}\Gamma_{5g}$ (see 
Sec.~\ref{sec:helmholtz}).
\bibitem{meaning} This would be the spectrum of the mean field hamiltonian 
${\cal T}(\vartheta,\phi)\langle{\cal T}(\vartheta,\phi)\rangle$ if we set 
$\langle{\cal T}(\vartheta,\phi)\rangle=1$.
\bibitem{scheme} While for general orientation, the spectrum of ${\cal T}$
consists of four singlets, the 111 directions are special by
possessing the level scheme singlet-doublet-singlet. We exploit
only the fact that the ground state is a singlet.
\bibitem{ehrenfest3} It means that (\ref{eq:ef}) is, though often
applicable, still system-specific, and not literally an Ehrenfest relation.
True Ehrenfest relations \cite{ehrenfest1} are thermodynamic identities, once
the assumption about finite discontinuities is made.
\bibitem{solterdos} A closely related observation was made by Solt and 
Erd\H os \cite{erdos,cf}. In their model, a ground state which appears 
non-magnetic if the field is applied in a certain direction, 
can be selected by quadrupolar interaction.
\bibitem{nonlin} In our present model, the vanishing of $a_H$ at the 
tricritical point is not generic; however, in some other multipolar models it
is. A detailed discussion of various possibilities admitted by the relation 
(\ref{eq:ef}) will be published elsewhere.
\bibitem{footnote} Though octupoles of other symmetries may be induced 
by the same fields.
\bibitem{Ising} There is a general resemblance to the well-known fact that the
ferromagnetic Ising model does not have a sharp phase transition
in an external magnetic field, basically because the field picks a
preferred spin direction, and thus there remains no symmetry to
break. More precisely: this rules out a continuous transition
only. Suitably constructed models may have a sharp first-order
transition. Similarly, our high-field analysis cannot exclude
first-order octupolar transitions. However, we found no evidence
for these.
\end{thebibliography}
\end{document}